\documentclass[epj,nopacs]{svjour}
\usepackage{epsfig,array}
\usepackage{multirow,color}
\newcommand{\be}{\begin{eqnarray}}
\newcommand{\ee}{\end{eqnarray}}
\newcommand{\bc}{\begin{center}}
\newcommand{\ec}{\end{center}}
\newcommand{\bea}{\begin{eqnarray}}
\newcommand{\eea}{\end{eqnarray}}

\newcommand{\beq}{\begin{equation}}
\newcommand{\eeq}{\end{equation}}

\newcommand{\nn}{\nonumber \\ }

\def\fun#1#2{\lower3.6pt\vbox{\baselineskip0pt\lineskip.9pt
\ialign{$\mathsurround=0pt#1\hfil##\hfil$\crcr#2\crcr\sim\crcr}}}

\begin{document}

\title{Photoproduction of pions and properties of baryon
resonances from a Bonn-Gatchina partial wave analysis}
\titlerunning{Photoproduction of pions and properties of baryon
resonances}
\author{
A.V.~Anisovich$\,^{1,2}$, E.~Klempt$\,^1$, V.A.~Nikonov$\,^{1,2}$,
M.A. Matveev$\,^{1,2}$, A.V.~Sarantsev$\,^{1,2}$, U.~Thoma$\,^{1}$ }
\authorrunning{A.V.~Anisovich \it et al.}
\institute{$^1\,$Helmholtz-Institut f\"ur Strahlen- und Kernphysik,
Universit\"at Bonn, Germany\\
$^2\,$Petersburg Nuclear Physics Institute, Gatchina, Russia}

\date{Received: \today / Revised version:}

\abstract{Masses, widths and photocouplings of baryon resonances are
determined in a coupled-channel partial wave analysis of a large
variety of data. The Bonn-Gatchina partial wave formalism is
extended to include a decomposition of t- and u-exchange amplitudes
into individual partial waves. The multipole transition amplitudes
for $\gamma p\to p\pi^0$ and $\gamma p\to n\pi^+$ are given and
compared to results from other analyses.
 \vspace{1mm}   \\
{\it PACS:  13.30.Eg, 13.40.Hq,  14.20.Gk}}

\maketitle

\section{Introduction}

The spectrum of baryon resonances is expected to be very rich and
considerably more complex than the mesonic excitation spectrum. Yet,
experimentally, the number of known light-quark mesons exceeds by
far the number of known baryon resonances \cite{Amsler:2008zzb}.

In quark models \cite{Capstick:bm,Glozman:1997ag,Loring:2001kx},
most high-mass baryon resonances are only weakly coupled to $N\pi$
\cite{Capstick:1992th}, and can thus not been seen in elastic $\pi
N$ scattering experiments. For inelastic reactions like $\pi N\to
\eta N$, $\pi N\to K\Lambda$, $\pi N\to K\Sigma$, data with a
polarized target are missing, and data on differential cross
sections have low statistics, and are often inconsistent when
different experiments are compared. States weakly coupled to the
$\pi N$ channel may thus have escaped identification. The situation
was aggravated by a recent analysis of a large body of $\pi N$
elastic and charge exchange scattering data in which many of the
less established nucleon and $\Delta$ resonances were not
confirmed~\cite{Arndt:2006bf}.

Other interpretations of the baryon spectrum exist as well. Very
popular are diquark models \cite{Anselmino:1992vg,Kirchbach:2001de,%
Jaffe:2003sg,Jaffe:2004ph} in which one quark-pair is frozen and in
which the number of predicted states decreases. Further, we mention
approaches based on chiral Lagrangians in which low-lying baryon
resonances are generated dynamically. In many cases, these
calculations offer a consistent description of resonance properties
and scattering data (see e.g. \cite{Kaiser:1995eg}) but so far, they
do not give a survey of all resonances to be expected. Often
discussed is the conjecture that chiral symmetry might be restored
in high-mass meson and baryon resonances.
\cite{Glozman:2003bt,Glozman:2007ek}. The conjecture gives an
attractive interpretation of one experimental observation, that
resonances show up as parity doublets or even higher multiplets. It
fails to predict at which mass resonances should be found and,
experimentally, all meson and baryon resonances on the leading Regge
trajectory have no parity partner. The AdS/QCD model describes QCD
in terms of a dual gravitational theory
\cite{deTeramond:2005su,Brodsky:2007hb}; with some phenomenological
adjustments, it is surprisingly successful in predicting the baryon
mass spectrum and the number of expected states
\cite{Forkel:2007cm,Forkel:2008un}. It also predicts where parity
multiplets should occur and where not. Reviews of baryon
spectroscopy can be found in
\cite{Hey:1982aj,Capstick:2000qj,Klempt:2009pi}.

A decision which of the above approaches provides the most accurate
representation of Nature requires a better experimental knowledge of
the excitation spectrum. The limitations of the current data base on
light-quark baryons is the stimulus for experiments studying baryon
resonances in photoproduction of complex final states where the $\pi
N$ channel can be avoided in both the initial and the final state.
The analysis of multibody final states including fermions is
complex; in order to compare different approaches and to identify
possible problems, a common meeting ground is needed. This is
provided by the simplest photoproduction reactions, by $\gamma p\to
p\pi^0$ and $\gamma p\to n\pi^+$. In this paper, we give masses,
widths, photocouplings, and $N\pi$ decay branching ratios for the
most important contributing resonances and compare our pion
photoproduction and helicity amplitudes to those obtained by SAID
\cite{SAID}, MAID \cite{MAID} and within the Gie\ss en model
\cite{Penner:2002md,Shklyar:2006xw}. The fits are based on a large
number of data sets and include data with multibody final states.
The study thus shows to which extent multibody final states are
compatible with the
best-studied $N\pi$ system.\\[-4ex]

\section{\label{Data}Data used in the fits}
A large number of reactions is used in the truly coupled-channel
fits presented here. The data cover elastic $\pi N$ scattering as
well as inelastic reactions, they cover differential cross sections
and single and double polarization variables. Reactions with
multi-body final states are included exploiting an event-based
likelihood method.

Different data sets often have a very different statistical power.
Weights $w_i$ are introduced to force the fit to take into account
highly significant but low-statistics data, e.g. beam asymmetries.
Without these weights, polarization data often have too small an
impact on the fit result. The weight of a newly introduced data set
is increased when the fit is visually unacceptable, or decreased
until first discrepancies between data and fit become apparent.

\subsection{Elastic \boldmath$\pi N\to \pi N$ scattering}
In the analysis presented here, data on elastic $\pi N$ scattering
(charge exchange is implicitly included) are not used directly.
Instead, we rely on the detailed work of the George-Washington
Center for Nuclear~Studies~\cite{Arndt:2006bf}~and~use, for energies
up to 2.2\,GeV, their scattering amplitudes.\\[-4ex]

\subsection{The reaction \boldmath$\pi^- p\to \eta n$}
The inelastic $\pi^- p$ scattering process leading to the $n\eta$
final state was reported from several experiments
\cite{Deinet:1969cd,Richards:1970cy,Brown:1979ii,Prakhov:2005qb} and
\cite{Debenham:1975bj,Crouch:1980vw}. Above 1.8\,GeV, large
discrepancies between the data \cite{Brown:1979ii,Crouch:1980vw}
show up. The data from \cite{Debenham:1975bj} cover extreme backward
angles and are partly incompatible with all other results. A
critical discussion of the available data can be found
in~\cite{Durand:2008es}. We use here the data from
\cite{Richards:1970cy,Prakhov:2005qb} (see Table
\ref{piN_data_table}) which show better consistency.

\begin{table}[t]
\caption{\label{piN_data_table}Pion induced reactions fitted in the
coupled-channel analysis and $\chi^2$ contributions.}
\bc\begin{tabular}{ccccc}
\hline\hline\\[-2ex]
$\pi N \rightarrow \pi N$& Wave & $N_{\rm data}$ &$w_i$ &$\chi^2/N_{\rm data}$\\[1ex]\hline\\[-2ex]
\cite{Arndt:2006bf}& $S_{11}$ & 104 & 30 & 1.81 \\
& $S_{31}$ & 112 & 20 & 2.27 \\
& $P_{11}$ & 112 & 20 & 2.49 \\
& $P_{31}$ & 104 & 20 & 2.01\\
& $P_{13}$ & 112 & 10 & 1.90 \\
& $P_{33}$ & 120 & 10 & 2.53\\
& $D_{13}$ & 96 & 10 & 2.16\\
& $D_{33}$ & 108 & 12 &2.56 \\
& $D_{15}$ & 96 & 20 & 3.37\\
& $F_{35}$ & 62 & 20 & 1.32\\
& $F_{37}$ & 72 & 10 & 2.86\\[0.5ex]\hline \hline\\[-2ex]
$\pi^- p \rightarrow \eta n$ & Observ. & $N_{\rm data}$&$w_i$ & $\chi^2/N_{\rm data}$\\[1ex]\hline\\[-2.3ex]
\cite{Richards:1970cy}&  $d\sigma/d\Omega$ & 70 & 10 &1.96 \\
\cite{Prakhov:2005qb}&  $d\sigma/d\Omega$ & 84 & 30 & 2.67 \\[0.5ex]\hline\\[-2.3ex]
\hline
\end{tabular}\vspace{-5mm}\ec
\end{table}

\begin{table}
\caption{\label{3BodyReactions}Reactions leading to 3-body final
states are included in event-based likelihood fits. The
$\chi^2/N_{\rm bin}$ values are calculated from selected Dalitz
plots (see text for details). References to the data are given in
the text.}
\bc\begin{tabular}{lccccc}
\hline\hline\\[-2ex]
\multicolumn{2}{c}{$d\sigma/d\Omega(\pi^-p \rightarrow \pi^0\pi^0
n)$} & $N_{\rm data}$ &$w_i$& $-\ln L$\\[1ex]\hline\\[-2ex]
T=373 MeV   && 5248 & 10& -1025\\
T=472 MeV   && 10641& 5& -2685\\
T=551 MeV   &\cite{Prakhov:2004zv}& 41172 & 2.5& -7322\\
T=655 MeV   && 63514 & 2& -15647\\
T=691 MeV   && 30030 & 3.5& -8256\\
T=733 MeV   && 29948 & 4& -7534\\[0.4ex]\hline\\[-2.1ex]
$d\sigma/d\Omega(\gamma p \rightarrow \pi^0\pi^0 p)$ &\cite{Thoma:2007bm,Sarantsev:2007bk}& 110601 & 4 & -27568\\
$d\sigma/d\Omega(\gamma p \rightarrow \pi^0\eta p)$
&\cite{Weinheimer:2003ng,Horn:2007pp,Horn:2008qv}& 17468 & 8 & -5587
\\ \hline\hline\\[-2ex]
\multicolumn{2}{c}{$d\sigma/d\Omega(\pi^-p \rightarrow \pi^0\pi^0
n)$} & $N_{\rm bin}$ & ~& $\chi^2/N_{\rm bin}$\\[1ex]\hline\\[-2ex]
T=373 MeV   && 471 & ~& 1.24\\
T=472 MeV   && 478& ~& 1.30\\
T=551 MeV   &\cite{Prakhov:2004zv}& 514 & ~& 1.56\\
T=655 MeV   && 518 & ~& 1.31\\
T=691 MeV   && 502 & ~ & 1.19\\
T=733 MeV   && 501 & ~& 1.53\\[0.4ex]\hline\\[-2.1ex]
$d\sigma/d\Omega(\gamma p \rightarrow \pi^0\pi^0 p)$ &\cite{Thoma:2007bm,Sarantsev:2007bk}& 769 & ~ & 1.59\\
$d\sigma/d\Omega(\gamma p \rightarrow \pi^0\eta p)$
&\cite{Weinheimer:2003ng,Horn:2007pp,Horn:2008qv}& 1119 & ~ & 1.04
\\
\hline\hline\\[-2ex]
\multicolumn{2}{c}{} & $N_{\rm data}$ &$w_i$& $\chi^2/N_{\rm data}$\\[1ex]\hline\\[-2ex]
$\Sigma(\gamma p \rightarrow \pi^0\pi^0 p)$ &\cite{Assafiri_03}
& 128 & 35 & 0.96\\
$\Sigma(\gamma p \rightarrow \pi^0\eta p)$ &\cite{Gutz:2008zz}& 180
& 15 & 2.37\\
 $E(\gamma p \rightarrow \pi^0\pi^0 p)$
&\cite{Ahrens_07} & 16 & 35 & 1.91
\\\hline\hline
\end{tabular}\vspace{-2mm}\ec
\end{table}

\subsection{The reaction \boldmath $\pi^- p \rightarrow \pi^0\pi^0 n$}
In the low-energy region, up to $\sim 1.5$\,GeV in mass, very
precise data from BNL are available \cite{Prakhov:2004zv}. These
data are included in an event-based likelihood fit (see Table
\ref{3BodyReactions}). The likelihood values have no direct
significance; only likelihood difference can be related to
probability changes when particular contributions are removed from
the fit. To demonstrate the quality of the description we have
constructed for every energy of the initial pion  $m^2_{p\pi^0}$
versus $m^2_{p\pi^0}$  Dalitz plots for data and for Monte Carlo
events with $40\times 40$ bins. The Monte Carlo events were weighted
with the squared amplitude from our final PWA solution. The data and
weighted Monte Carlo Dalitz plots were compared; in
Table~\ref{3BodyReactions} the $\chi^2/N_{bin}$ is given as well as
the number of bins with nonzero number of Monte Carlo events.

\begin{table}[t]
\caption{\label{chisquare}Observables from $\pi$ and $\eta$
photoproduction fitted in the coupled-channel analysis and $\chi^2$
contributions. For pion production, free normalization factors and
additional systematic errors were introduced to allow for data
variation beyond statistical expectations (see text).}
\bc\begin{tabular}{lcccc}
\hline\hline\\[-2ex]
$\gamma p \rightarrow \pi^0 p$ & Observ. & $N_{\rm data}$&$w_i$ & $\chi^2/N_{\rm data}$\\[1ex]\hline\\[-2ex]
\cite{Fuchs:1996ja} (TAPS@MAMI)& $d\sigma/d\Omega$  & 1692 & 1.5&1.25 \\
\cite{Ahrens:2002gu,Ahrens:2004pf} (GDH A2)& $d\sigma/d\Omega$  & 164 & 7&1.34 \\
\cite{Bartalini:2005wx} (GRAAL)& $d\sigma/d\Omega$  & 861 & 2&1.46 \\
\cite{Bartholomy:2004uz,vanPee:2007tw} (CB-ELSA)& $d\sigma/d\Omega$  & 1106 & 3.5&1.34 \\
\cite{Dugger:2007bt} (CLAS)& $d\sigma/d\Omega$ & 592 & 5 &2.11 \\
\cite{Bartalini:2005wx,Barbiellini:1970qu,Gorbenko:1974sz,Gorbenko:1978re,Belyaev:1983xf,%
Blanpied:1992nn,Beck:1997ew,Adamian:2000yi,Blanpied:2001ae}&  $\Sigma$ &1492& 3 &3.26\\
\cite{Gorbenko:1974sz,Gorbenko:1978re,Belyaev:1983xf,Booth:1976es,Feller:1976ta,%
Gorbenko:1977rd,Herr:1977vx,Fukushima:1977xj,Bussey:1979wt,Agababian:1989kd,%
Asaturian:1986bj,Bock:1998rk,Maloy:1961qy}&  $T$&389& 6 &3.71\\
\cite{Gorbenko:1974sz,%
Gorbenko:1978re,Belyaev:1983xf,Maloy:1961qy,Gorbenko:1975pz,Kato:1979br,Bratashevsky:1980dk,%
Bratashevsky:1986xz}&  $P$&607& 3 &3.23\\
\cite{Bussey:1979wr,Ahrens:2005zq} &  $G$&75& 5 &1.50\\
\cite{Bussey:1979wr} &  $H$&71& 5 &1.26\\
\cite{Ahrens:2002gu,Ahrens:2004pf} &  $E$&140& 7 &1.23\\
\cite{Bratashevsky:1980dk,Avakyan:1991pj}&  $O_x$&7& 10 &1.77\\
\cite{Bratashevsky:1980dk,Avakyan:1991pj}&  $O_z$&7& 10 &0.46\\\hline\\[-2.3ex]
\hline\\[-2ex]
$\gamma p \rightarrow \pi^+ n$ & Observ. & $N_{\rm data}$&$w_i$ & $\chi^2/N_{\rm data}$\\[1ex]\hline\\[-2ex]
\cite{Ecklund:1967zz,Betourne:1968bd,Bouquet:1971cv,Fujii:1971qe,%
Ekstrand:1972rt,Fujii:1976jg,Arai:1977kb,Durwen:1980mq,Althoff:1983te,%
Heise:1988ag,Buechler:1994jg,Dannhausen:2001yz,Ahrens:2006gp}&  $d\sigma/d\Omega$ & 1583 & 2 &1.64  \\
\cite{Ahrens:2004pf,Ahrens:2006gp} (GDH A2)&  $d\sigma/d\Omega$ & 408 & 14 &0.61  \\
\cite{Dugger:2009pn} (CLAS)&  $d\sigma/d\Omega$ & 484 & 4 &1.80  \\
\cite{Blanpied:2001ae,Taylor:1960dn,Smith:1963zza,Alspector:1972pw,Knies:1974zx,%
Ganenko:1976rf,Bussey:1979ju,Getman:1981qt,Hampe:1980jb,Beck:1999ge,%
Ajaka:2000rj,Bocquet:2001ny}&  $\Sigma$ &899 & 3 &3.48\\
\cite{Bussey:1979ju,Getman:1981qt,Althoff:1973kb,Arai:1973xs,Feller:1974qf,Althoff:1975kt,Genzel:1975tx,%
Althoff:1976gq,Althoff:1977ef,Fukushima:1977xh,Getman:1980pw,%
Fujii:1981kx,Dutz:1996uc}&  $T$&661 & 3 &3.21\\
\cite{Bussey:1979ju,Getman:1981qt,Egawa:1981uj}&  $P$&252 & 3 &2.90\\
\cite{Ahrens:2005zq,Bussey:1980fb,Belyaev:1985sp} &  $G$&86 & 3 &5.64\\
\cite{Bussey:1980fb,Belyaev:1985sp,Belyaev:1986va} &  $H$&128& 3& 3.90\\
\cite{Ahrens:2004pf,Ahrens:2006gp} &  $E$&231& 14 & 1.55\\\hline\\[-2.3ex]
\hline\\[-2ex]
$\gamma p \rightarrow \eta p$ & Observ. & $N_{\rm data}$&$w_i$ & $\chi^2/N_{\rm data}$\\[1ex]\hline\\[-2ex]
\cite{Krusche:nv}& $d\sigma/d\Omega$ &100 & 7 &2.16 \\
\cite{Crede:04,Bartholomy:2007zz}& $d\sigma/d\Omega$ &680 & 40 &1.47 \\
\cite{Ajaka:1998zi}& $\Sigma$ &51 & 10 &2.26 \\
\cite{Bartalini:2007fg} &  $\Sigma$ &100& 15 &2.02\\
\cite{Bock:1998rk} &  $T$ &50& 70 &1.48\\
\hline\\[-2ex]
\hline\\[-2ex]
\end{tabular}\vspace{-2mm}\ec
\end{table}

\subsection{Photoproduction of single neutral pions off protons}

References to the data on the reaction $\gamma p\to p\pi^0$ and
their $\chi^2$ contributions are collected in Table~\ref{chisquare}.
For the differential cross section, we use only the most recent
data, reported by TAPS@MAMI \cite{Fuchs:1996ja}, GDH-A2
\cite{Ahrens:2002gu,Ahrens:2004pf}, GRAAL \cite{Bartalini:2005wx},
CB-ELSA \cite{Bartholomy:2004uz,vanPee:2007tw}, and CLAS
\cite{Dugger:2007bt} which cover a wide range of energies and
angles. A large variety of older data exist which cover only a
limited fraction of the energy and angular range. These data provide
significant information on polarization observables.

\begin{figure}
\centerline{
\epsfig{file=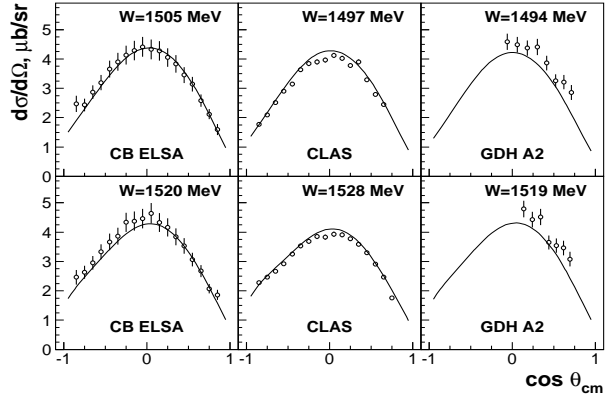,width=0.45\textwidth,height=0.3\textwidth}
} \caption{The CB-ELSA, CLAS  and GDH differential cross section on
$\gamma p\to \pi^0 p$ in the region 1500 MeV. In the Crystal Barrel
data, a common systematic error due to uncertainties in the
reconstruction efficiency is included. The curve represents
the``first" fit without normalization (see text). The GDH data are
introduced with a large weight.}
\label{fig:pi0_cb_cl_gdh}       
\end{figure}

The differential cross sections reported by the different
collaborations exhibit small but significant systematic
discrepancies practically in all mass regions; due to the small
statistical errors these are easily recognized. We show the
systematic deviations by comparing the data with a curve
representing the ``first" fit to all data, without normalization
factors.

In Fig.~\ref{fig:pi0_cb_cl_gdh} the differential cross section from
GDH-A2, CLAS and CB-ELSA are shown and compared to the preliminary
fit for the 1495-1530\,MeV mass range. The GDH data systematically
exceed CLAS data while the CB-ELSA data provide numbers between
these two measurements. The GDH-A2 data have a larger statistical
error than the CLAS data; we introduced them into the fit with
larger weight since they provide important information about the
difference between helicity 3/2 and 1/2 cross sections.

\begin{figure}
\centerline{
\epsfig{file=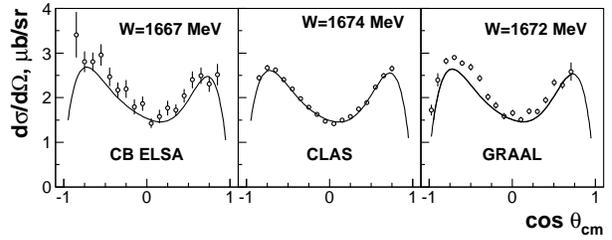,width=0.45\textwidth} }
\caption{Comparison of three data sets on the $\gamma p\to \pi^0 p$
differential cross section in the region 1670 MeV. The curve
represents the ``first" fit without normalization (see text).}
\label{fig:diff_pi0}       
\end{figure}

In the mass region 1600-1750 MeV, there are notable discrepancies
between GRAAL and CLAS data (here the CB-ELSA data fall again
between GRAAL and CLAS results). As an example, the mass region
around 1670 MeV is shown for the three data sets in
Fig.~\ref{fig:diff_pi0}. The curve corresponds to the fit where the
CLAS data are taken with statistical errors only and dominate the
solution.

\begin{figure}
\bc
\epsfig{file=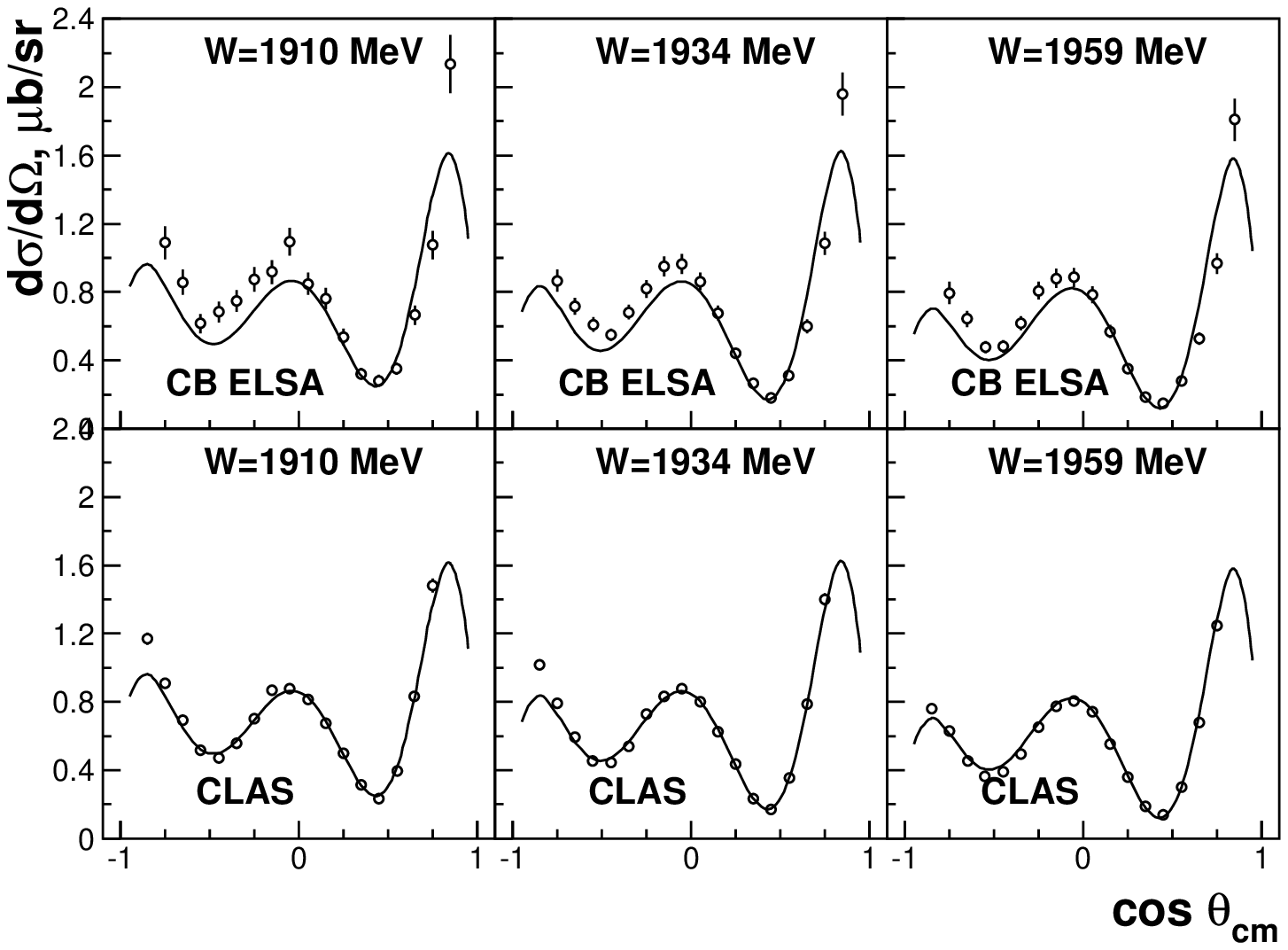,width=0.45\textwidth,height=0.28\textwidth}\\
\epsfig{file=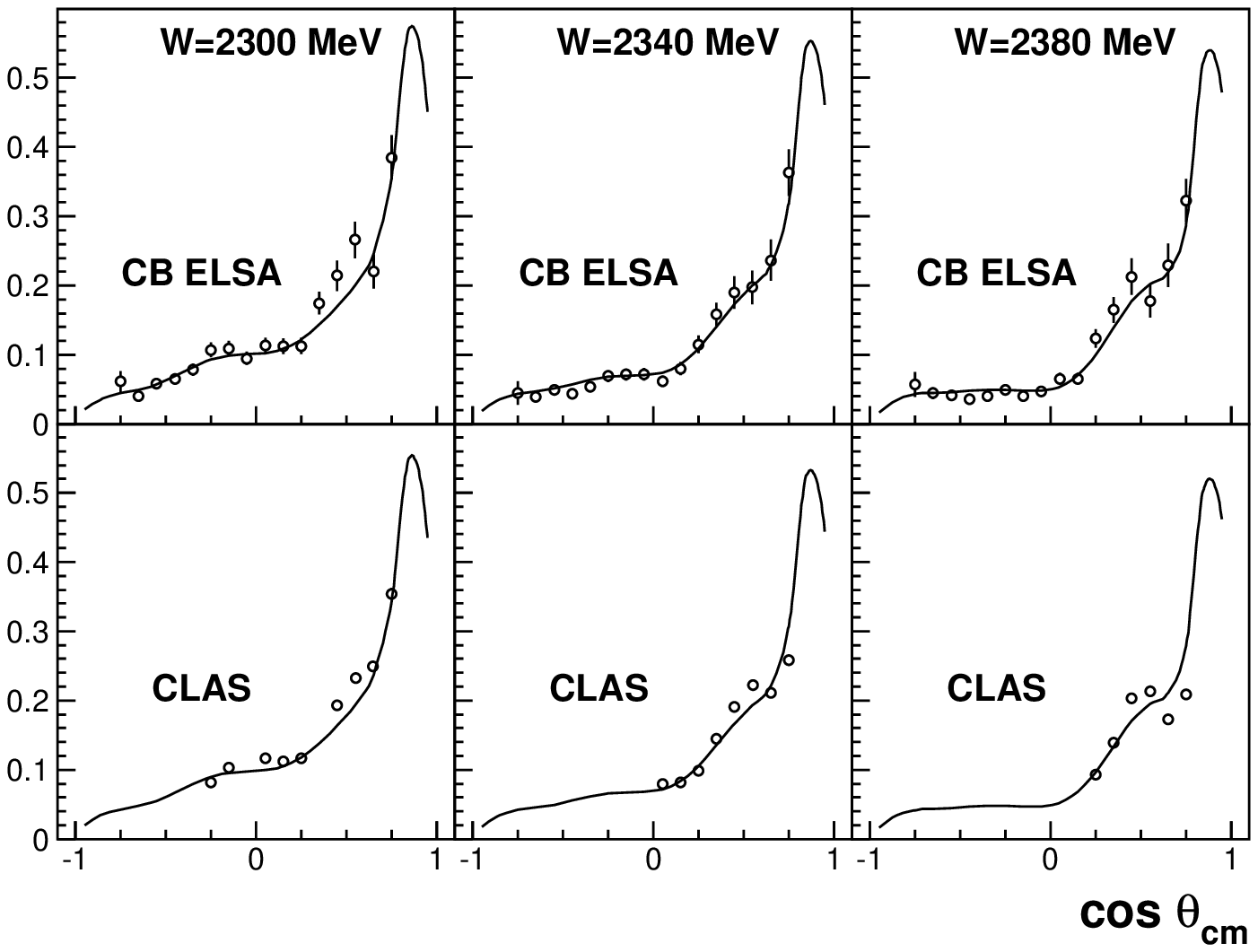,width=0.45\textwidth,height=0.28\textwidth}
\vspace{-2mm}\ec \caption{The CB-ELSA and CLAS differential cross
section on $\gamma p\to \pi^0 p$ in the region of 1900 (top) and
2350 (bottom) MeV. The curve represents the ``first" fit without
normalization (see text).\vspace{-2mm}}
\label{fig:pi0_cb_cl_1900}       
\end{figure}

At higher energies, only the CB-ELSA and CLAS data are available.
There are two clear discrepancies between these data sets. The first
one is located in the 1900\,MeV mass region where the CB-ELSA data
systematically exceed the CLAS data in the backward hemisphere (see
Fig.~\ref{fig:pi0_cb_cl_1900}, top). A second discrepancy shows up
above $W=2100$ MeV in the very forward angular range. Here the
corresponding CB-ELSA points are systematically lower than those
from CLAS. At $W$=2300-2400\,MeV, the two data sets are fully
consistent (see Fig.~\ref{fig:pi0_cb_cl_1900}, bottom).

A fraction of the discrepancies is assigned to normalization. Most
experiments give explicit normalization errors which were not taken
into account in the ``first" fit. We then allowed  for a free
normalization factor for the $\pi^0$ differential cross section,
which is determined in the fit to 1.00 (TAPS@MAMI), 1.01 (GDH-A2),
0.99 (GRAAL and CB-ELSA) and 0.95 (CLAS).

The data from the four experiments are still not yet statistically
compatible; at least one experiment must have additional
unrecognized systematic errors. Of course, we do not know which
experiment. We assume that all four experiments have systematic
errors which were not recognized. These were estimated from the
variance of the experimental results in preset bins of energy and
angle (in $\cos\theta$). For this purpose, differential cross
sections were calculated, by interpolation, for these bins. From the
variance we estimated systematic errors which increase linearly from
1\% at $W=1400$\,MeV to 9\% at $W=2450$\,MeV. These systematic
errors were added to all four data sets. With the increased
systematic errors, the data are compatible and the $\chi^2$ of a fit
reflects the quality of a fit and not the inconsistency between
different data sets. These errors are used in the fits only. In the
figures, the data and their errors are shown as quoted in the
original papers.

The beam asymmetry $\Sigma$ has been determined in a number of
experiments
\cite{Bartalini:2005wx,Barbiellini:1970qu,Gorbenko:1974sz,Gorbenko:1978re,Belyaev:1983xf,%
Blanpied:1992nn,Beck:1997ew,Adamian:2000yi,Blanpied:2001ae}, as well
as the target asymmetry $T$
\cite{Gorbenko:1974sz,Gorbenko:1978re,Belyaev:1983xf,Booth:1976es,Feller:1976ta,%
Gorbenko:1977rd,Herr:1977vx,Fukushima:1977xj,Bussey:1979wt,Agababian:1989kd,%
Asaturian:1986bj,Bock:1998rk}, and
the polarization $P$ of the recoiling proton \cite{Gorbenko:1974sz,%
Gorbenko:1978re,Belyaev:1983xf,Maloy:1961qy,Gorbenko:1975pz,Kato:1979br,Bratashevsky:1980dk,%
Bratashevsky:1986xz}. Few data exist from experiments with polarized
photons and polarized target or from measurements of the recoil
polarization. Data on $O_x$ and $O_z$ can be found in
\cite{Bratashevsky:1980dk,Avakyan:1991pj}, on $G$ in
\cite{Bussey:1979wr,Ahrens:2005zq}, and on $H$ in
\cite{Bussey:1979wr}. Data on the helicity difference $\sigma_{3/2}-
\sigma_{1/2}$ were published in \cite{Ahrens:2002gu,Ahrens:2004pf};
in Tables \ref{3BodyReactions} and \ref{chisquare} we quote $E$
which is defined as $(\sigma_{3/2}- \sigma_{1/2})/(\sigma_{3/2}+
\sigma_{1/2})$. These data are included in the fits. Their
statistical errors are mostly large, the systematic errors likely
small. Hence we retain the original errors.

\subsection{The reaction \boldmath$\gamma p \rightarrow \pi^+ n$}

\begin{figure}
\centerline{
\epsfig{file=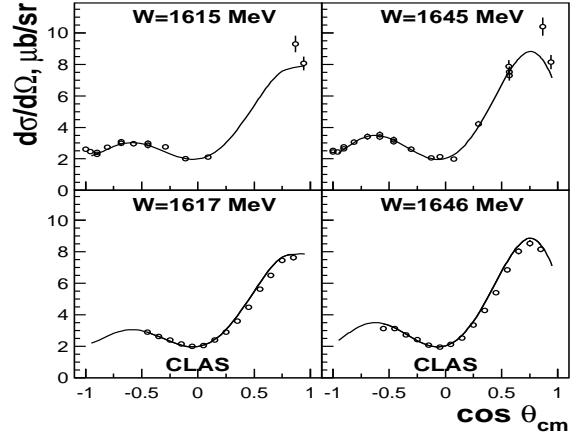,width=0.43\textwidth,height=0.33\textwidth}
} \caption{Older data from different experiments (top) and CLAS
differential cross section on $\gamma p\to \pi^+ n$ in the region
1630 MeV. The curve represents the ``first" fit without
normalization (see text). The consistency is excellent.}
\label{fig:cl_said_pipn}       
\end{figure}

Total cross sections were reported in
\cite{Ecklund:1967zz,Betourne:1968bd,Bouquet:1971cv,Fujii:1971qe,%
Ekstrand:1972rt,Fujii:1976jg,Arai:1977kb,Durwen:1980mq,Althoff:1983te,%
Heise:1988ag,Zenz:1988ah,Buechler:1994jg,Dannhausen:2001yz,Ahrens:2006gp,%
Dugger:2009pn}. Again, some discrepancies show up, at energies above
1600\,MeV and in the forward region, between the new CLAS data and
former measurements. An example of such discrepancies is given in
Fig.~\ref{fig:cl_said_pipn}. Normalization factors for the different
data were introduced which are determined to be in the range from
0.96 to 1.03. A consistent description was achieved by adding a
systematic error which increases linearly from 1\% at $W=1400$\,MeV
and to 9\% at $W=2450$\,MeV.

The beam asymmetry $\Sigma$ was determined in
\cite{Blanpied:2001ae,Taylor:1960dn,Smith:1963zza,Alspector:1972pw,Knies:1974zx,%
Ganenko:1976rf,Bussey:1979ju,Getman:1981qt,Hampe:1980jb,Beck:1999ge,%
Ajaka:2000rj,Bocquet:2001ny}, the target asymmetry $T$ in
\cite{Bussey:1979ju,Getman:1981qt,Althoff:1973kb,Arai:1973xs,Feller:1974qf,Althoff:1975kt,Genzel:1975tx,%
Althoff:1976gq,Althoff:1977ef,Fukushima:1977xh,Getman:1980pw,%
Fujii:1981kx,Dutz:1996uc}, the neutron recoil polarization can be
found in \cite{Bussey:1979ju,Getman:1981qt,Egawa:1981uj}. A few data
from double polarization are available: on $G$
\cite{Ahrens:2005zq,Bussey:1980fb,Belyaev:1985sp}, $H$
\cite{Bussey:1980fb,Belyaev:1985sp,Belyaev:1986va}, and on the
helicity difference $\sigma_{3/2}- \sigma_{1/2}$
\cite{Ahrens:2004pf,Ahrens:2006gp}. The data are fitted with the
errors as given in the respective papers.

\subsection{Photoproduction of \boldmath$\eta$ mesons off protons}

For photoproduction of $\eta$ mesons, differential cross sections
\cite{Krusche:nv,Dugger:2002ft,Crede:04,Bartholomy:2007zz,Bartalini:2007fg}
and the related beam asymmetry $\Sigma$
\cite{Ajaka:1998zi,Bartalini:2007fg,Elsner:2007hm} are the only
quantities which have been measured so far. Double polarization
observables are presently studied intensively at several
laboratories but so far, no results have been published. The recent
high-statistics measurements on $\gamma p\to p\eta$
\cite{Williams:2009yj,Crede:2009zz} are not yet included in the fits
presented here.
\begin{table}[t]
\caption{\label{chisquare1}Hyperon photoproduction observables
fitted in the coupled-channel analysis and $\chi^2$ contributions.}
\bc\begin{tabular}{lcccc}
\hline\hline\\[-2ex]
$\gamma p \rightarrow K^+ \Lambda$ & Observ. & $N_{\rm data}$&$w_i$ & $\chi^2/N_{\rm data}$\\[1ex]\hline\\[-2ex]
\cite{Bradford:2005pt} &  $d\sigma/d\Omega$&1377 & 4 &1.81 \\
\cite{Zegers:2003ux}&  $\Sigma$ &45& 10 & 1.65\\
\cite{Lleres:2007tx}&  $\Sigma$ &66& 5 & 1.53\\
\cite{McNabb:2003nf} &  $P$&202 & 6.5 &2.03\\
\cite{Lleres:2007tx} &  $P$&66 & 3 &1.26\\
\cite{Lleres:2008em}&  $T$&66 & 15 &1.26\\
\cite{Bradford:2006ba} &  $C_x$&160 &11 &1.23\\
\cite{Bradford:2006ba}&  $C_z$&160 & 11 &1.41\\
\cite{Lleres:2008em}&  $O_x$&66 & 12 &1.30\\
\cite{Lleres:2008em}&  $O_z$&66 & 15 &1.54\\\hline\\[-2.3ex]
\hline\\[-2ex]
$\gamma p \rightarrow K^+ \Sigma$ & Observ. & $N_{\rm data}$&$w_i$ & $\chi^2/N_{\rm data}$\\[1ex]\hline\\[-2ex]
\cite{Bradford:2005pt}& $d\sigma/d\Omega$ & 1280& 2.5 &2.06 \\
\cite{Zegers:2003ux} &  $\Sigma$ &45& 10 &1.11\\
\cite{Lleres:2007tx} &  $\Sigma$ &42& 5 &0.90\\
\cite{McNabb:2003nf}&  $P$&95 & 6 &1.45\\
\cite{Bradford:2006ba}&  $C_x$&94 & 7 &2.20\\
\cite{Bradford:2006ba}&  $C_z$&94 & 7 &2.00\\\hline\\[-2.3ex]
\hline\\[-2ex]
$\gamma p \rightarrow K^0 \Sigma^+$ & Obsv. & $N_{\rm data}$&$w_i$ & $\chi^2/N_{\rm data}$\\[1ex]\hline\\[-2ex]
\cite{McNabb:2003nf} &  $d\sigma/d\Omega$ & 48 &2.3 &3.76 \\
\cite{Lawall:2005np} &  $d\sigma/d\Omega$ & 160 &5 &0.98 \\
\cite{Castelijns:2007qt} &  $d\sigma/d\Omega$ & 72 &5 &0.82 \\
\cite{Castelijns:2007qt}&  $P$&72 & 20 &0.61\\
\hline\hline
\end{tabular}\ec
\end{table}

\subsection{The reactions \boldmath$\gamma p\to K^+\Lambda, K^+\Sigma^0$ and $K^0\Sigma^+$}

Data on hyperon photoproduction used in the present fits are
collected in Table~\ref{chisquare1}. We use the differential cross
sections for $\gamma p\to K^+\Lambda$ and $K^+\Sigma^0$ from CLAS
\cite{Bradford:2005pt}. As shown in \cite{Sarantsev:2005tg}, the
Saphir data \cite{Glander:2003jw} on differential cross sections are
about compatible with the CLAS data when an energy dependent
normalization factor is introduced. The beam asymmetry was measured
at SPring-8 \cite{Zegers:2003ux} and GRAAL \cite{Lleres:2007tx}; the
$\Lambda$ polarization was deduced in \cite{Lleres:2007tx} and
\cite{McNabb:2003nf}. Target asymmetry $T$ and $O_x$ and $O_z$ for
$\gamma p\to K^+\Lambda$ were reported in \cite{Lleres:2008em}. In
\cite{Bradford:2006ba}, CLAS data on the spin transfer coefficients
$C_x$ and $C_z$ were presented for both, $\gamma p\to K^+\Lambda$
and $\gamma p\to K^+\Sigma^0$.

Differential cross sections on the reaction $\gamma p \rightarrow
K^0 \Sigma^+$ were measured by CLAS \cite{McNabb:2003nf}, Saphir
\cite{Lawall:2005np} and CB-ELSA/ TAPS \cite{Castelijns:2007qt}. For
the latter data we include the determination of the $P$ polarization
derived from an analysis of the $\Sigma^+$ decay.

\subsection{The reactions \boldmath$\gamma p\to p\pi^0\pi^0$ and $\gamma p\to p\pi^0\eta$}

The two reactions $\gamma p\to p\pi^0\pi^0$
\cite{Thoma:2007bm,Sarantsev:2007bk} and $\gamma p\to p\pi^0\eta$
\cite{Weinheimer:2003ng,Horn:2007pp,Horn:2008qv} are included event
by event using an extended likelihood method. The quality of the fit
can be judged from the description of Dalits plots. For the $\gamma
p\to p\pi^0\pi^0$ reaction we constructed Dalitz plots in
$m^2_{p\pi^0}$ versus $m^2_{p\pi^0}$ with $20\times 20$ bins, for
the four 100 MeV $\gamma p$ invariant mass intervals from 1350 to
1750 MeV. For the $\gamma p\to p\eta\pi^0$ reaction, $m^2_{p\pi^0}$
versus $m^2_{p\eta}$ Dalitz plots were constructed for seven 100 MeV
$\gamma p$ invariant mass intervals from 1700 to 2400 MeV. The
number of bins with nonzero Monte Carlo events and $\chi^2/N_{bin}$
are given in Table~\ref{3BodyReactions}. The beam asymmetries
\cite{Assafiri_03,Gutz:2008zz} and the helicity dependence $E$
\cite{Ahrens_07} are included in the fit in the form of histograms.

Both these reactions have been studied intensively, see
\cite{Braghieri_95,Haerter_97,Zabrodin_97,Zabrodin_99,Wolf_00,%
Kleber_00,Langgaertner_01,Ripani_03,Ahrens_03,Kotulla_04,%
Ahrens_05,Strauch_05,Ajaka_07,Krambrich:2009te} for the first and
\cite{Nakabayashi:2006ut,Ajaka:2008zz,Kashevarov:2009ww} for the
latter reaction. For these data only selected histograms are
available; they are not included in our fits.

\section{Partial wave amplitudes}

A general expression for the decomposition of the two-particle
scattering amplitude $A(s,t)$ into partial wave amplitudes
$A^{\beta\beta'}_n(s)$ which describe production, propagation and
decay of a two-particle systems with fixed total angular momentum
$J$, parity  and (if conserved) $C$-parity can be written as:
\be
A(s,t)&\!=&\!\!\sum\limits_{\beta\beta' n}\!\! A^{\beta\beta'}_n(s)
Q^{(\beta) \dagger}_{\mu_1\ldots\mu_n}(k)
F^{\mu_1\ldots\mu_n}_{\nu_1\ldots\nu_n}
Q^{(\beta')}_{\nu_1\ldots\nu_n}(q)~~~~
\label{decomp_0}
\ee
where $k_i$ are initial and $q_i$ are final particle momenta,
$s=(k_1+k_2)=(q_1+q_2)=P^2$, $t=(k_1-q_1)^2=(k_2-q_2)^2$,
$k=(k_1-k_2)/2$, $q=(q_1-q_2)/2$ and $n=J$ for a boson system and
$n=J-1/2$ for a fermion one. The vertices
$Q^{(\beta')}_{\nu_1\ldots\nu_n}$ and
$Q^{(\beta)\dagger}_{\mu_1\ldots\mu_n}$ ('$\dagger$' stands for
hermitian conjugation) describe the transition of the system into
the initial- and final-state particles, and depend on the total and
relative momenta. The indices $\beta$ and $\beta'$ list quantum
numbers of the production and decay amplitudes, e.g. isospin, spin
and orbital angular momenta. The tensor
$F^{\mu_1\ldots\mu_n}_{\nu_1\ldots\nu_n}$ depends only on the total
momentum $P$ and describes the tensor structure of the partial wave.
It is often called projection operator. The formalism for
construction of vertices for meson-baryon partial waves and
projection operators is given in
\cite{Anisovich:2004zz,Anisovich:2006bc}. For convenience we provide
key formula for projection operators and vertices in Appendix A.

In the case of resonance production, the total amplitude $A(s,t)$
can be expanded into a sum of partial wave amplitudes multiplied by
vertices, see eq.~(\ref{decomp_0}). Here the partial wave amplitudes
$A^{\beta\beta'}_n(s)$ provide the energy dependence of the
resonance which can be parameterized, for example, as $N/D$
amplitude, as K-matrix or, in the simplest case, as a Breit-Wigner
amplitude \cite{Anisovich:2008zz}. For non-resonant contributions,
like $t$ and $u$ channel exchanges, the situation is different. In
many partial wave analyses (including the present one) these
contributions are simply added to the resonant part of the total
amplitude and the sum is used to fit the experimental data. However,
one needs to know the contribution of $t$ and $u$-exchanges in every
partial wave if the final partial wave amplitudes are to be compared
with results from other analyses. This decomposition is also
required when rescattering between non-resonant and resonant parts
of the amplitude should be taken into account. For the non-resonant
contributions used in the energy dependent fits one has therefore to
solve an inverse task: to extract partial wave amplitudes from the
total amplitude.

This task can be solved by using the orthogonality condition for
partial wave operators. Multiplying the total amplitude from eq.
(\ref{decomp_0}) with initial and final projection operators and
vertices and integrating over solid angle of the initial and final
momenta we obtain
\be
\label{decomp_2}
&&F^{\tau_1\ldots\tau_n}_{\mu_1\ldots\mu_n}\int
\frac{d\Omega_k}{4\pi}\frac{d\Omega_q}{4\pi}
Q^{(\alpha)}_{\mu_1\ldots\mu_n}(k) A(s,t)
Q^{(\alpha')}_{\nu_1\ldots\nu_n}(q)
F^{\nu_1\ldots\nu_n}_{\eta_1\ldots\eta_n}
\nn &&
= (-1)^n
F^{\tau_1\ldots\tau_n}_{\eta_1\ldots\eta_n}
\sum \limits_{\beta\beta'}
A^{\beta\beta'}_n(s)W^{\alpha\beta}_n(k^2_\perp)
W^{\beta'\alpha'}_n(q^2_\perp)\,,
\ee
where $k^2_\perp$ and $q^2_\perp$ are squared relative momenta
orthogonal to the total momentum of the system $P$ (see Appendix A).

The factor $W_n^{\alpha\beta}$ corresponds to the on-shell one-loop
amplitude for transition between two vertices
$Q^{(\beta)}_{\mu_1\ldots\mu_n}$. It can be calculated as
\be
W^{\alpha\beta}_n(k^2_\perp)\!&=&\!
\frac{F^{\alpha_1\ldots\alpha_n}_{\mu_1\ldots\mu_n}}{\xi_n}
\!\!\int\!\!
\frac{d\Omega_k}{4\pi}Q^{(\alpha)}_{\mu_1\ldots\mu_{n}}(k)
Q^{(\beta)}_{\nu_1\ldots\nu_{n}}(k)F_{\alpha_1\ldots\alpha_n}^{\nu_1\ldots\nu_n}
\nn
\xi_n&=&(-1)^nF^{\nu_1\ldots\nu_n}_{\mu_1\ldots\mu_n}g_{\mu_1\nu_1}\ldots
g_{\mu_n\nu_n}\,.
\label{wn}
\ee
For meson-nucleon and $\gamma N$ vertices, the $W^{\alpha\beta}_n$
were calculated in \cite{Anisovich:2006bc}. For convenience we
provide the corresponding expressions in Appendix B and expressions
for partial wave amplitudes for photoproduction of a single meson
are given in Appendix C.

\subsection{Parameterization of the partial wave amplitudes}

In the present analysis,  the partial waves at low energies are
described in the framework of a K-matrix/P-vector approach.
High-mass resonances (above 2.2 GeV) are described by relativistic
multi-channel Breit-Wigner amplitudes. In the case of
photoproduction reactions, the regge\-ized t- and u-channel
amplitudes were added to the resonant part. Then the multipoles were
calculated by solving eq.~(\ref{decomp_2}).

\subsubsection{Pion induced reactions in K-matrix approach}

The multi-channel amplitude is given by the matrix $\hat
\textbf{A}(s)$ where the matrix element ${A}_{ab}(s)$ defines the
transition amplitude from state 'a' to state 'b'. In
eq.~(\ref{decomp_0}) this amplitude is denoted as
$A^{\beta\beta'}_n(s)$ to emphasize the different spin-parity
contributions. Now we will use the notation ${A}_{ab}(s)$ which
identifies the initial and the final channels, e.g. $\gamma N$, $\pi
N$, $\eta N$, $K\Lambda$, $\pi \Delta$, and omit the indices
representing the partial wave. Scattering between different channels
is taken into account explicitly in the K-matrix; the amplitude is
given by
 \be
 \mathbf{\hat A}(s) \;=\; \mathbf{\hat K}\;(\mathbf{\hat I}\;-\;i
 \mathbf{\hat \rho \hat K})^{-1}\,. \label{k_matrix}
\ee
where $\mathbf{\hat K}$ is the K-matrix, $\mathbf{\hat I}$ is the
unity matrix and $\mathbf{\hat \rho}$ is a diagonal matrix of the
according phase space. For two-particle states (for example $\pi
N$), the phase space is calculated as a simple loop diagram (see
\cite{Anisovich:2006bc}). For $J=L+1/2$, the so-called '+' states,
the phase space is equal to
\be
\label{psr_plus}
\rho_+(s)=\frac{\alpha_L}{2L+1} \frac{2|\vec
k|^{2L+1}}{\sqrt{s}}\frac{k_{10}+m_N}{2m_N} \frac
{F(k^2)}{B(L,r,k^2)}
\ee
and for '-' states with $J=L-1/2$, the phase space is given by
\be
\label{psr_minus}
\rho_{-}(s)=\frac{\alpha_L}{L} \frac{2|\vec
k|^{2L+1}}{\sqrt{s}}\frac{k_{10}+m_N}{2m_N} \frac
{F(k^2)}{B(L,r,k^2)}
\ee
where $s$ is the total energy squared, $k$, is the relative momentum
between baryon and meson, $\vec k$ its three-vector component,
$k_{10}$ is the energy of the baryon (with mass $m_N$) calculated in
the c.m.s. of the reaction. $J$ is the total, $L$ the orbital
angular momentum of the baryon-plus-meson system, and the
coefficient $\alpha_L$ is equal to:
\be
\alpha_L=\prod\limits_{n=1}^L\frac{2n-1}{n}\,.
\ee
The phase volume is regularized at large energies by a standard
Blatt-Weisskopf form $B(L,r,k^2)$ with $r=0.8$\,fm, and a
form-factor $F(k^2)$ of the type
 \be
F(k^2)=\frac{\Lambda+0.5}{\Lambda+k^2}\quad{\rm or}\quad
F(k^2)=\frac{\Lambda+2.5}{\Lambda+s}\,.
 \ee
Fits with both parameterizations yield nearly identical results. The
parameter $\Lambda$ were taken from our previous analysis
\cite{Sarantsev:2005tg,Anisovich:2005tf} and fixed to 1.5 for the
first parameterization and 3.0 for the second one. The exact
formulas for the three-body phase volume are given in
\cite{Anisovich:2006bc}.

The K-matrix $\mathbf{\hat K}$ is parameterized as follows:
 \be
 K_{ab}\;=\;\sum_\alpha \frac{g_a^{(\alpha)} g_b^{(\alpha)}} {M^2_\alpha - s} \;+\; f_{ab},
 \label{Kmat}
 \ee
where $M_\alpha$ and $g_a^{(\alpha)}$ are the mass and the coupling
constant of the resonance $\alpha$, and where $f_{ab}$ describes a
direct (non-resonant) transition from the initial state $a$ to the
final state $b$, e.g. from $\pi N\to\Lambda K$.

For most partial waves it is sufficient to assume that $f_{ab}$ are
constants. The $S_{11}$ and $S_{31}$ and waves require a slightly
more complicated structure, we use
\be
 f_{ab} =\frac{f_{ab}^{(1)}+f_{ab}^{(2)}\sqrt s}{s-s_0^{ab}}\,.
\ee
 Here the $f_{ab}^{(i)}$ and $s_0^{ab}$ are constants
which are determined in the fits. In the case of the $S_{11}$ wave,
this more flexible parameterization is required to describe $\pi
N\to N\pi$, $\pi N\to N\eta$, and $\eta N\to N\eta$ transitions. Let
us note that this form is similar to the one used by SAID
\cite{Arndt:2006bf}.

\subsubsection{The photoproduction amplitude}

The photoproduction amplitude can be written in the P-vector
approach \cite{Chung:1995dx}. The P-vector amplitude for the initial
state '$a$' photoproduction is then given by
 \be
  A_a \;=\; \hat P_b\;(\hat I\;-\;i\hat \rho \hat K)^{-1}_{ba}\,.
\ee
The production  vector $\mathbf{\hat P}$ is parameterized as:
 \be
 P_{b}\;=\;\sum_\alpha \frac{ g_{\gamma \rm N}^{(\alpha)} g_b^{(\alpha)}}{M^2_\alpha - s} \;+\;
 \tilde f_{b}
 \label{Pvect}
 \ee
where $g_{\gamma\rm N}^{(\alpha)}$ are the photo-couplings of the
resonance $\alpha$ and where non-resonant production of a final
state $b$ is described by contributions $\tilde f_{b}$. In general,
these are functions of $s$ but mostly, a constant $\tilde f_{b}$ is
sufficient.

The P-vector approach is based on the idea that a channel with a
weak coupling can be omitted from the K-matrix. Indeed, adding to
the K-matrix the $\gamma N$ channel would not change  the properties
of the amplitude. Due to its weak coupling, the $\gamma N$
interaction can be taken into account only once; this is done in the
form of a P-vector. Loops due to virtual decays of a resonance into
$N\gamma$ and back into the resonance can be neglected safely. A
similar approach can be used to describe decay modes with a weak
couplings. The amplitude for the transition into such a channel can
be written as D-vector amplitude,
\be
\label{amplitude}  A_a \;=\; \hat D_a + [\hat K (\hat I\;-\;i\hat
\rho \hat K)^{-1}\,\hat \rho ]_{ab} \hat D_{b}\;,
\ee
 where the parameterization of the
D-vector is similar to the parameterization of the P-vector:
\be
 D_{b}\;=\;\sum_\alpha \frac{g_b^{(\alpha)}g_{f}^{(\alpha)} }{M^2_\alpha - s} \;+\;
 \tilde d_{b}\,.
 \label{Dvect}
 \ee
Here $g_{f}^{(\alpha)}$ is the coupling of a resonance to the final
state and $\tilde d_{b}$ is a non-resonant production from the
K-matrix-channel $b$ to the final state. As in the case of the
P-vector approach, channels with weak couplings can be taken into
account only in their final decay, and are not taken into account in
the rescattering. Let us note that if the final state is already
included as one of K-matrix channels, the amplitude
(\ref{amplitude}) reproduces the K-matrix amplitude
(\ref{k_matrix}).

In cases where both, initial and final coupling constants are weak,
we use an approximation which we call PD-vector. In this case the
amplitude is given by
\be
A_{ab} \;=\; \hat G_{ab} + \hat P_{a}
(\hat I\;-\;i\hat \rho \hat K)^{-1}\,\hat \rho \hat D_{b}\;,
\ee
where $\hat G_{ab}$ corresponds to a tree diagram for the transition
from state '$a$' to state '$b$'.
\be
 G_{ab}\;=\;\sum_\alpha \frac{g_a^{(\alpha)}g_{b}^{(\alpha)} }{M^2_\alpha - s} \;+\;
 \tilde h_{ab}\,.
 \label{PDvect}
 \ee
Here $g_{i}^{(\alpha)}$ is the production coupling of the resonance.
For photoproduction, $g_{a}^{(\alpha)}=g_{\gamma\rm N}^{(\alpha)}$
holds true, and $\tilde h_{ab}$ is the direct non-resonant
transition from the initial to the different final channels.

\subsection {Reggeized meson  exchange amplitudes}

At high energies, angular distributions of photo-produced mesons
exhibit clear peaks in the forward direction. These peaks originate
from meson exchanges in the t-channel. Their contributions are
parameterized as $\pi$, $\rho(\omega)$, $\rm K$ or $\rm K^*$
exchanges.

The most straight forward parameterization of particle exchange
amplitudes is the exchange of Regge trajectories. The invariant part
of the t-channel exchange amplitude can be written as
\cite{Anisovich:2008zz}

\be
T(s,t)=g_1(t)g_2(t) R(\pm,\nu,t)\,\;\;\;\; \nu=\frac 12 (s-u).
\ee
Here, $g_i$ are vertex functions, and $R(+,\nu,t)$ and $R(-,\nu,t)$
are Reggeon propagators for exchanges with positive and negative
signature. Exchanges of $\pi$ and $\rm K$ have positive, $\rho$,
$\omega$ and $\rm K^*$ exchanges have negative signature.

The $\rho$ trajectory has a negative signature and the corresponding
propagator is equal to
\be
R_\rho(-,\nu,t)=\frac{ie^{-i\frac{\pi}{2}\alpha_\rho(t)}} {\cos
(\frac{\pi}{2}\alpha_\rho(t)) \Gamma \left (\frac
{\alpha_\rho(t)}{2} +\frac 12\right )} \left
(\frac{\nu}{\nu_0}\right )^{\alpha_\rho(t)} \ .\quad
\ee
where $\alpha_{\rho}(t)=0.50+0.85t$. The $\omega$ trajectory is
identical to the $\rho$ trajectory. The expressions for other
Reggeon propagators used in the fit are given in Appendix D.

\section{Partial wave analysis}
\subsection{Fit of the $\pi^0 p$ and $\pi^+n$ photoproduction reactions}

\begin{figure}[pt]
\centerline{
\epsfig{file=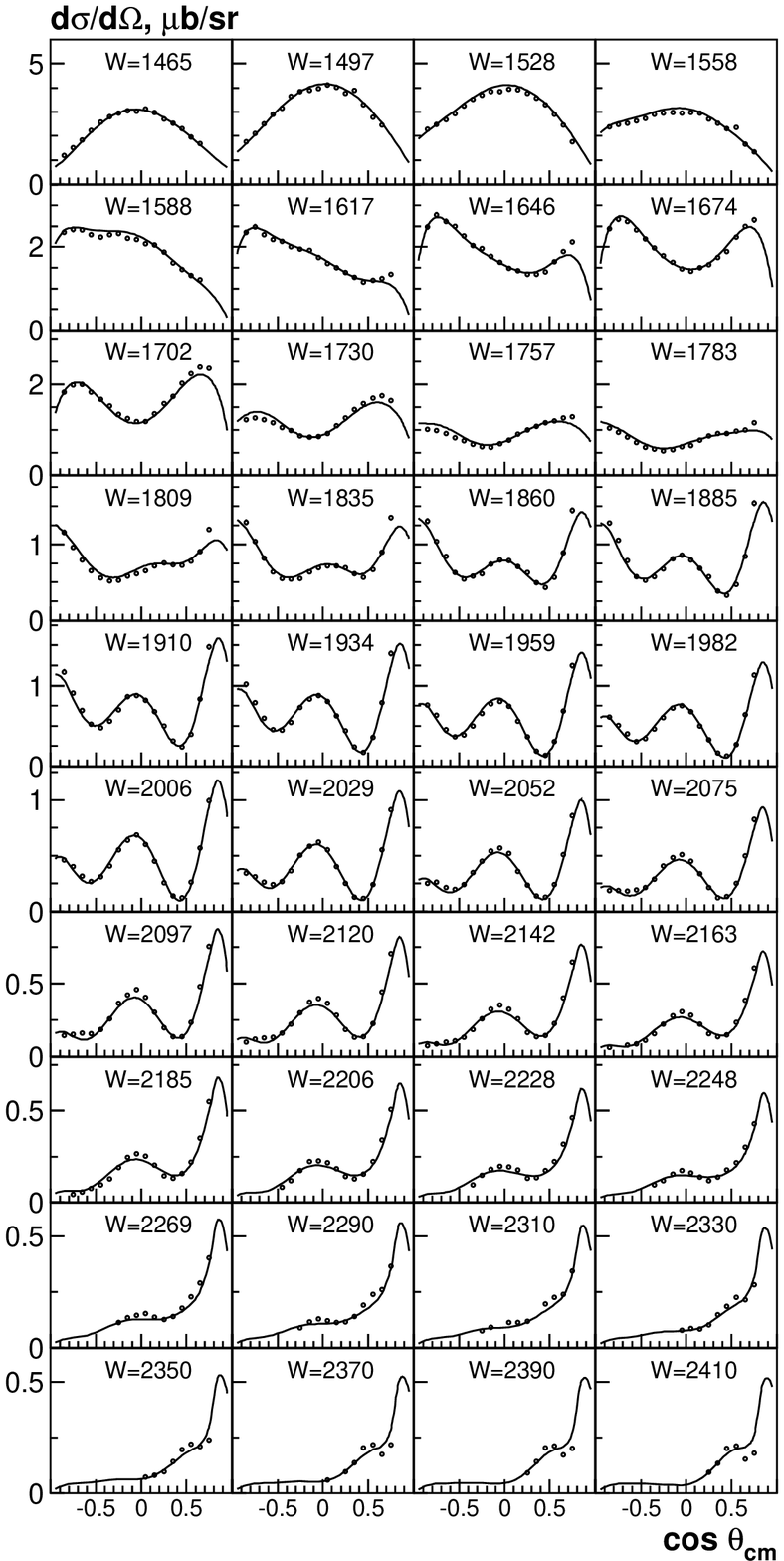,width=0.5\textwidth,height=0.66\textheight}
} \caption{CLAS data on the differential cross section for $\gamma
p\to \pi^0p$ with current solution. Only statistical errors for the
CLAS data are shown}\vspace{4mm}
\label{fig:pi0_clas}       
\centerline{
\epsfig{file=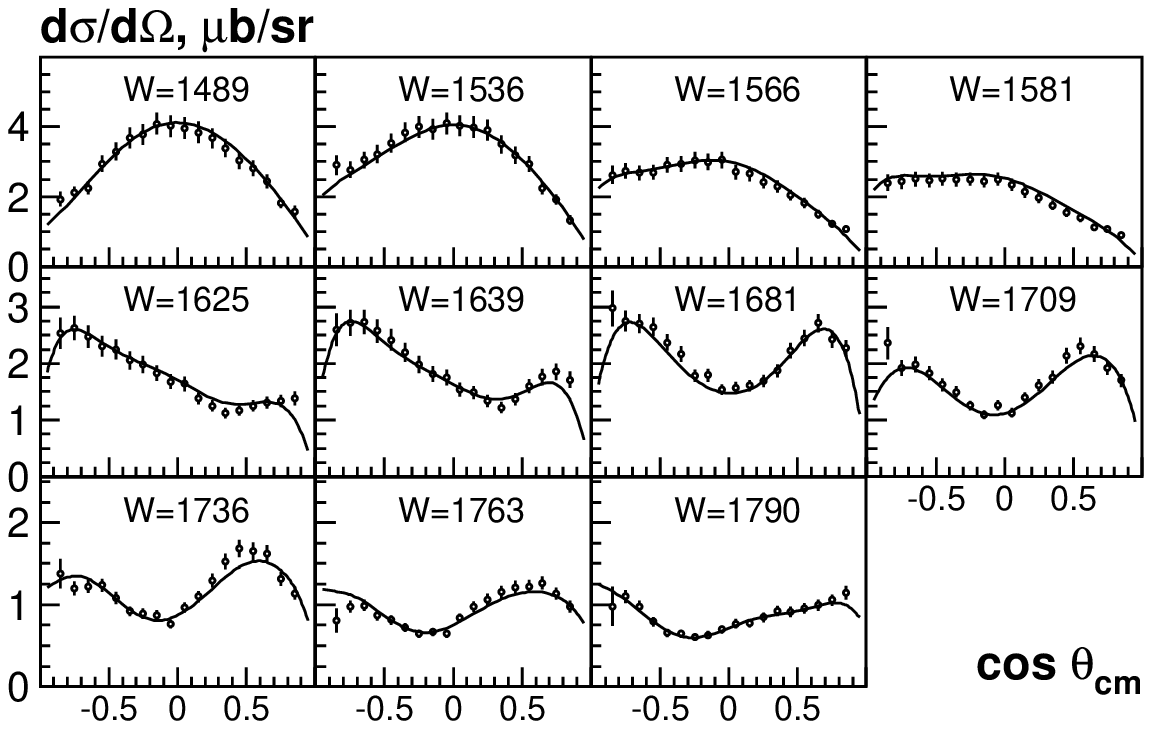,width=0.50\textwidth,height=0.22\textheight}
} \caption{CB-ELSA data on the $\gamma p\to \pi^0p$ differential
cross section for energies below 1.8 GeV. Only the errors quoted by
the CB-ELSA collaboration are shown.\vspace{3mm}}
\label{fig:pi0_cb}       
\end{figure}

The new CLAS data on the $\gamma p\to \pi^0p$ reaction are compared
to our fit in Fig.~\ref{fig:pi0_clas}. The $\chi^2$ contributions of
this fit from the various channels are given in
Tables~\ref{piN_data_table}\,-\,\ref{chisquare1}. We remind the
reader that we estimated additional systematic errors for the
$\gamma p\to \pi^0p$ and $\gamma p\to \pi^+n$ differential cross
sections; these additional errors are not shown in
Fig.~\ref{fig:pi0_clas}.

\begin{figure}
\centerline{
\hspace{5mm}\epsfig{file=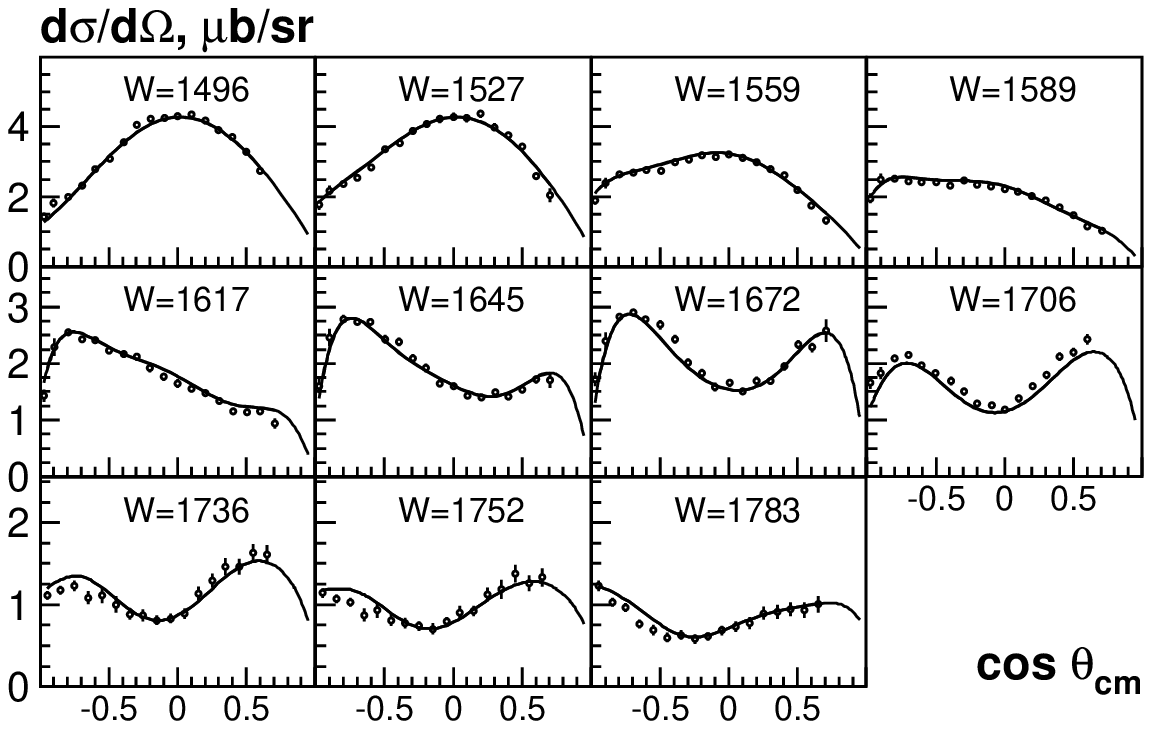,width=0.48\textwidth} }
\caption{GRAAL data on the $\gamma p\to \pi^0p$ differential cross
section at energies close to the CLAS values. Only the errors quoted
by the GRAAL collaboration are shown.\vspace{5mm}}
\label{fig:pi0_graal}       
\centerline{
\epsfig{file=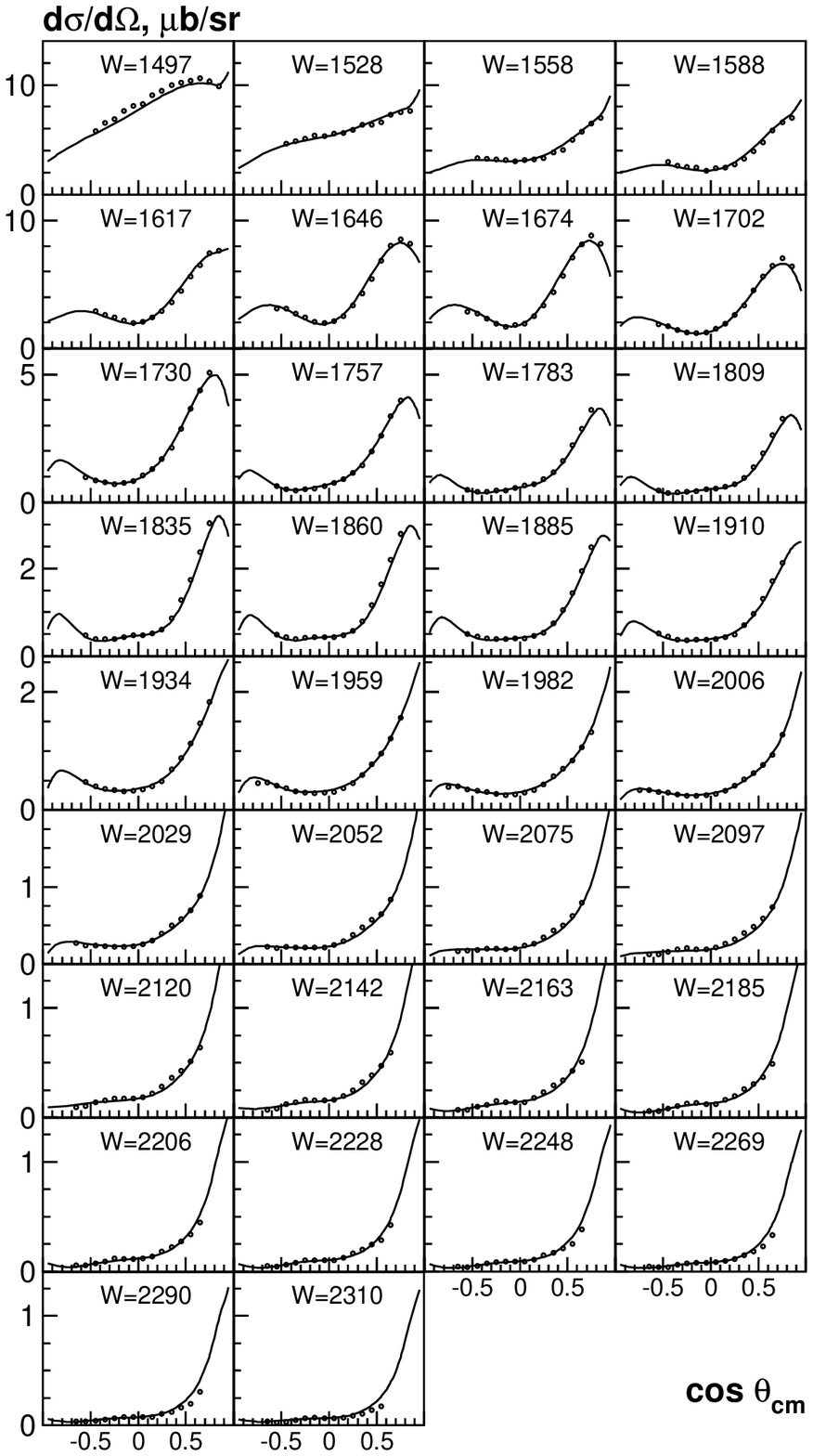,width=0.50\textwidth,height=0.6\textheight}
} \caption{CLAS data on the $\gamma p\to \pi^+n$ differential cross
section with the current solution. \vspace{4mm}}
\label{fig:pipn_clas}       
\end{figure}

Some systematic deviations between data and fit can be recognized in
the mass region below 1800\,MeV. These are mostly the result of
discrepancies between the data. In Fig.~\ref{fig:pi0_cb} we show for
comparison some CB-ELSA data, and in Fig.~\ref{fig:pi0_graal} some
GRAAL data, for invariant masses which are close to the CLAS values.
At higher energies the solution describes very well the new CLAS
data, however CB-ELSA data are also described with rather good
accuracy.

\begin{figure}[pt]
\centerline{
\epsfig{file=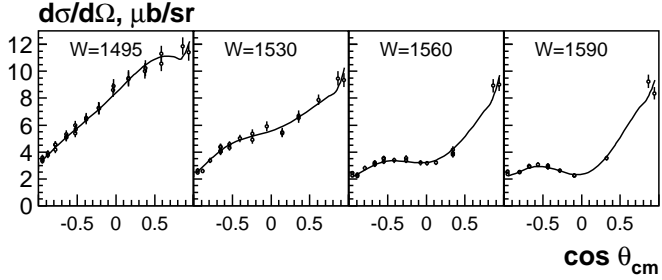,width=0.50\textwidth} }
\caption{The $\gamma p\to \pi^+n$ differential cross section with
the current solution. The data are taken from
\cite{Bouquet:1971cv,Ekstrand:1972rt,Fujii:1976jg,Arai:1977kb,Durwen:1980mq,Althoff:1983te,Heise:1988ag,Dannhausen:2001yz}}
\label{fig:pipn_said}       
\end{figure}

The new data on the $\gamma p\to \pi^+n$ and the fit curve are shown
in Fig.~\ref{fig:pipn_clas}. Here, the total normalization factors
resolve rather well discrepancies  at masses below 1600 MeV. The
description of the earlier data on the $\gamma p\to \pi^+n$ in this
mass region is shown in Fig.~\ref{fig:pipn_said}.

\subsection{Photoproduction multipoles}

We now turn to a discussion of the partial wave amplitudes. It
should be stressed that the amplitudes we give for $\gamma p \to
p\pi^0$ and $\gamma p \to n\pi^+$ are constrained by a large number
of other reactions. This is particularly important in the vicinity
of thresholds. Of course, the elastic $\pi N$ scattering amplitude
and the pion photoproduction amplitude are influenced by opening new
channels and the couplings to the new channels can be estimated from
their effect on the scattering and photoproduction amplitudes. But
this is rather indirect, and it is desirable to take the inelastic
channels into account directly.

The multipoles for $\pi^0$ photoproduction are shown in
Fig.~\ref{fig:pi0_mult_re} in comparison to the SAID SP09K2700
\cite{SAID} and
MAID 2007 \cite{MAID} solutions, those for  $\gamma p\to \pi^+n$ in
Fig.~\ref{fig:pipn_mult_re}. The errors cover a large number of fits
which differ mostly by the parameterization of the $2\pi N$ channel
at masses above 1.8\,GeV. For convenience of the reader, we list in
Table \ref{waves} the lowest photoproduction multipoles and the
corresponding partial waves.

\begin{figure*}[pt]
\centerline{
\epsfig{file=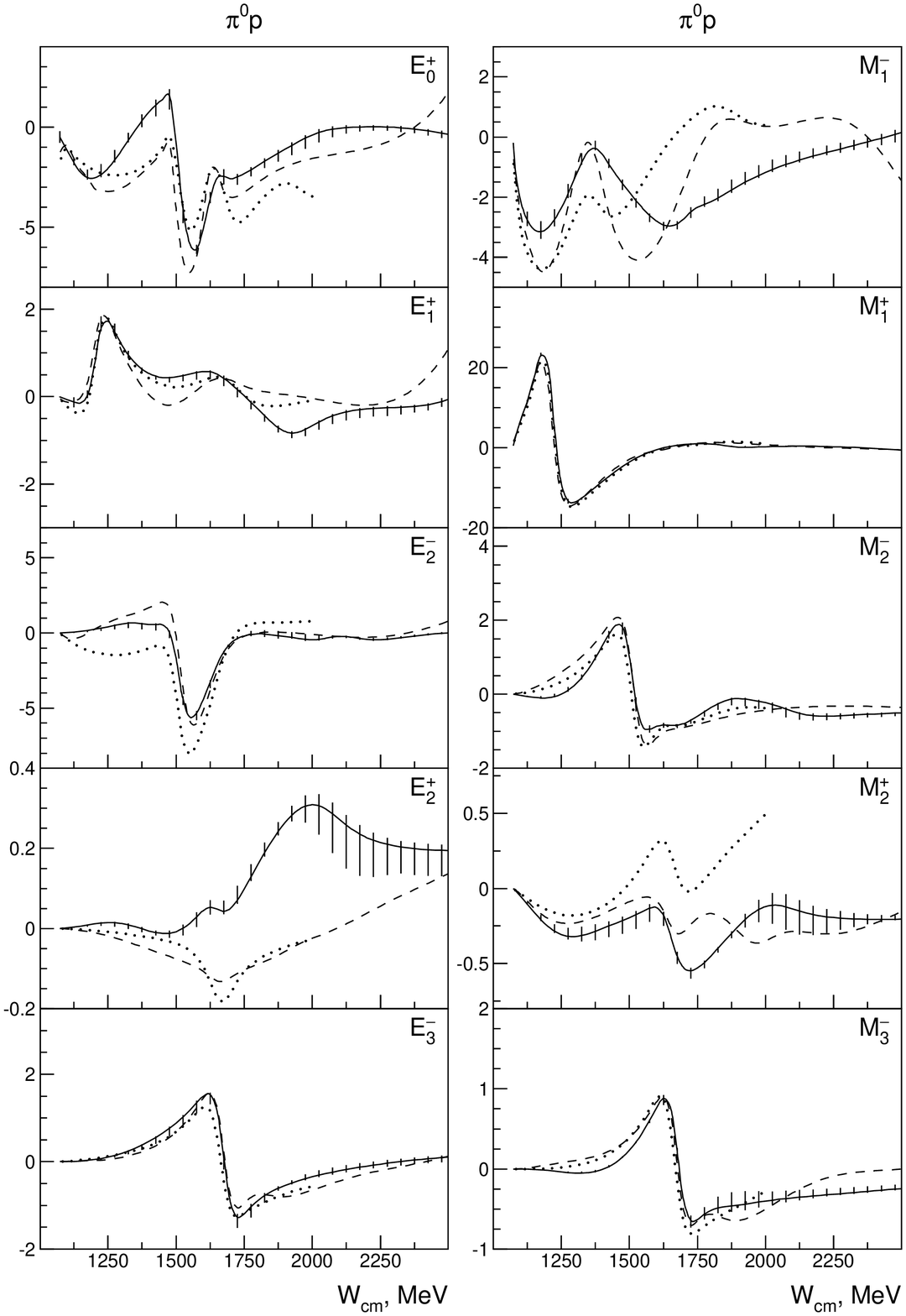,width=0.46\textwidth,height=0.45\textheight}~~
\epsfig{file=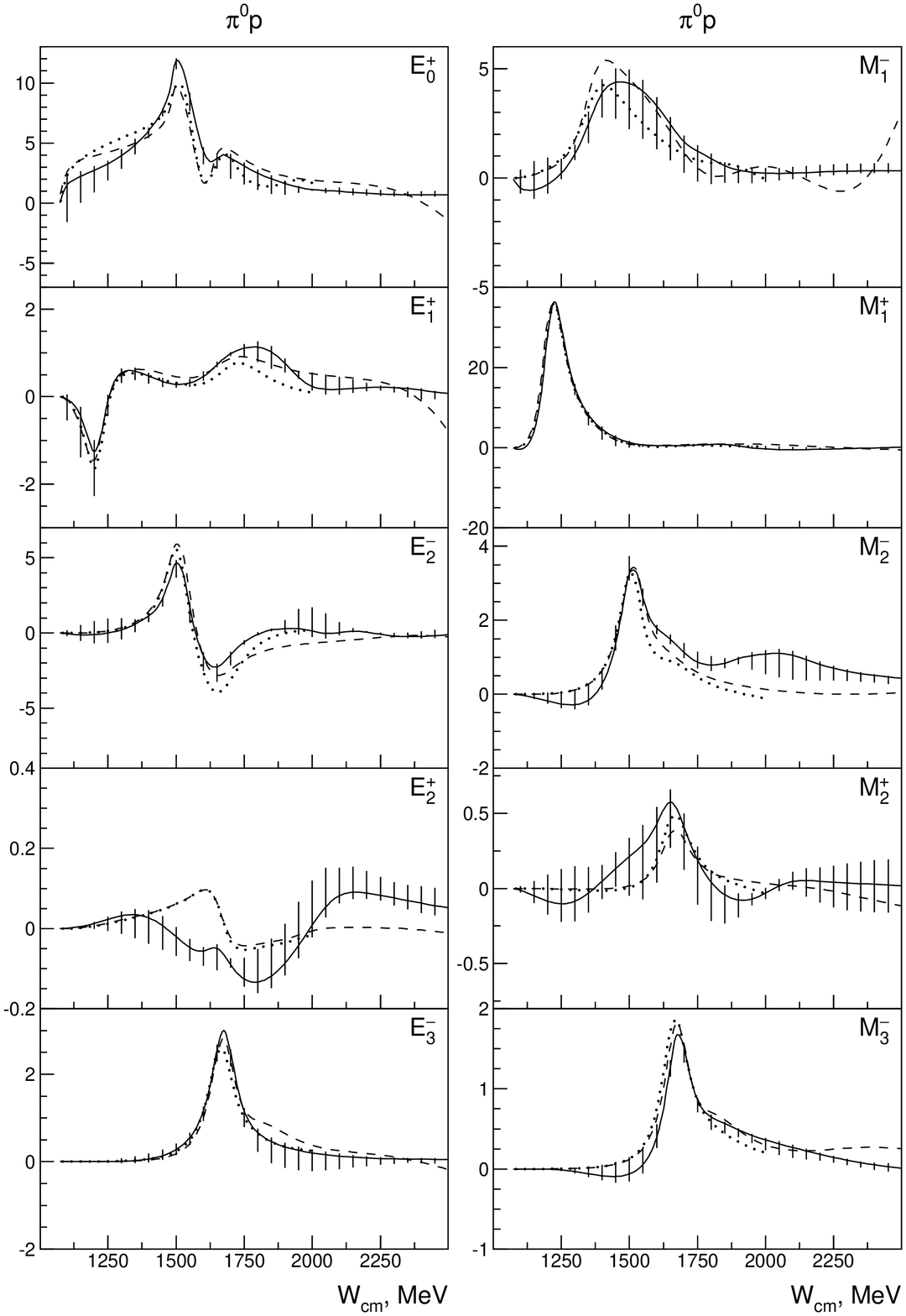,width=0.46\textwidth,height=0.45\textheight}~~
} \caption{\label{fig:pi0_mult_re} The real (two left-hand columns)
and imaginary (two right-hand columns) part of multipoles for the
$\pi^0$ photoproduction. The errors are systematic and cover a large
number of fits (see the text). The dashed curves correspond to the
SAID solution SP09K2700 \cite{SAID} and the dotted curves to the
MAID solution 2007 \cite{MAID}} \centerline{
\epsfig{file=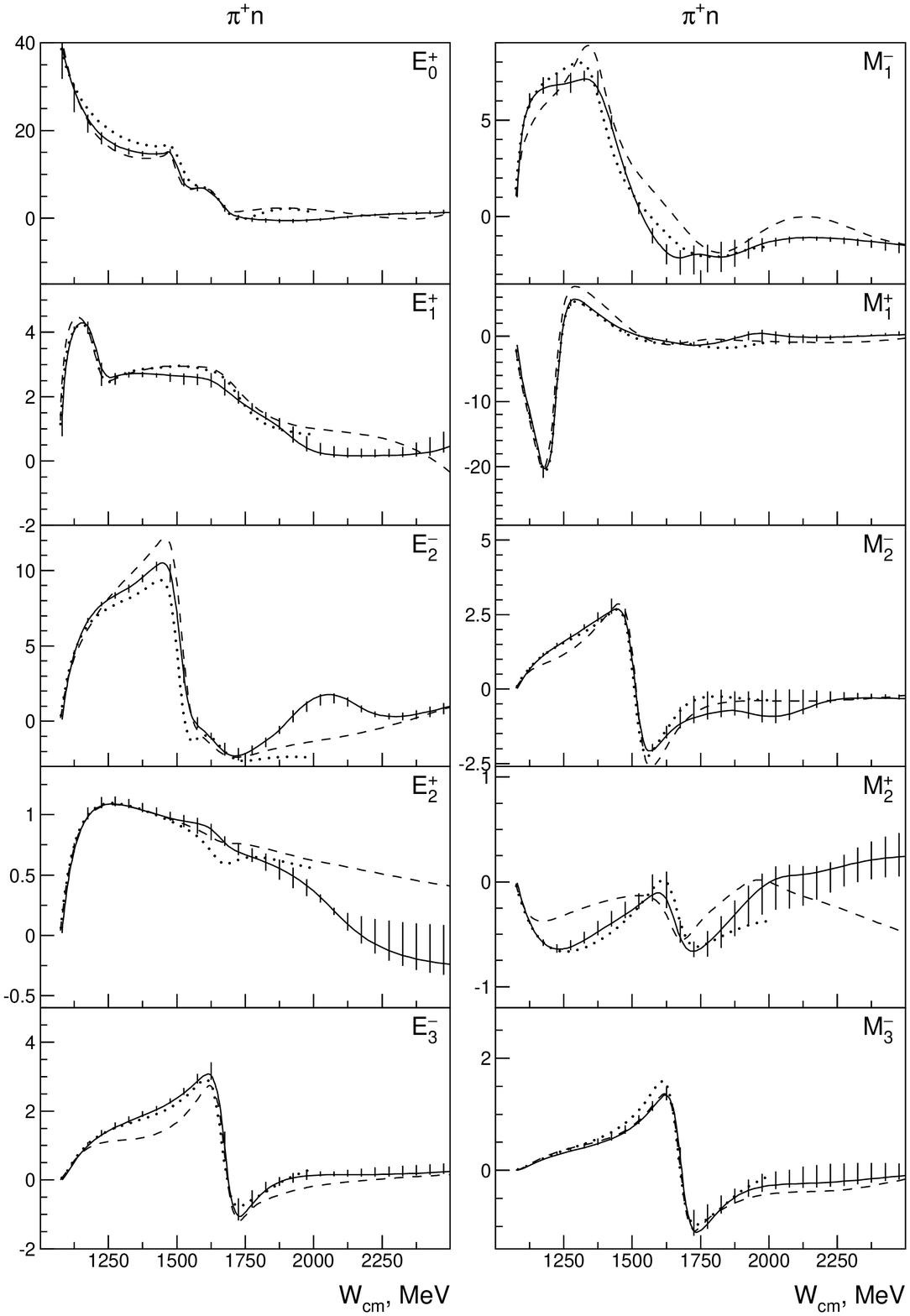,width=0.46\textwidth,height=0.45\textheight}~~
\epsfig{file=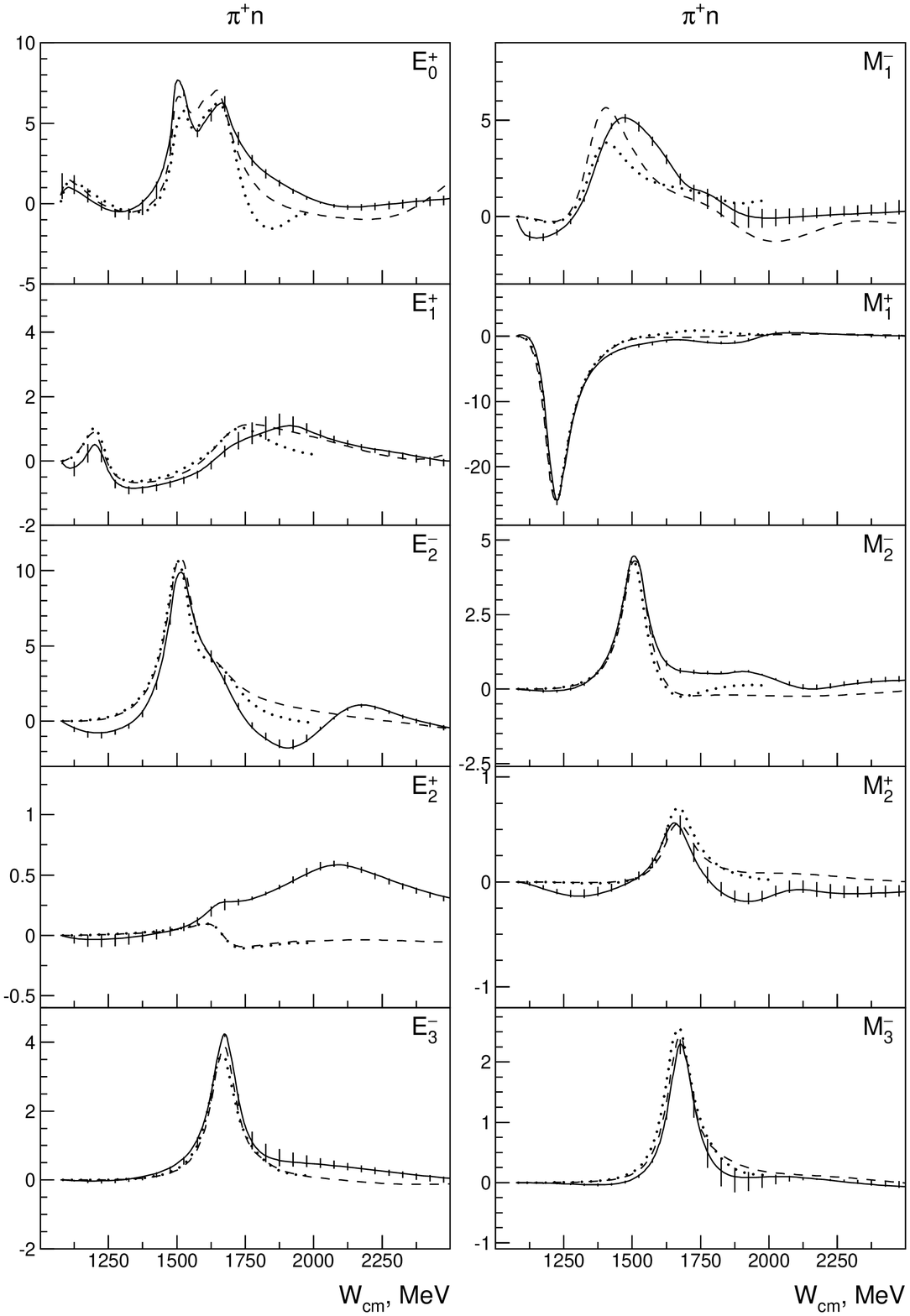,width=0.46\textwidth,height=0.45\textheight}~~
} \caption{\label{fig:pipn_mult_re} The real (two left-hand columns)
and imaginary (two right-hand columns) part of multipoles for the
$\gamma p\to \pi^+n$ reaction. The errors are systematic and cover a
large number of fits (see the text). The dashed curves correspond to
the SAID solution SP09K2700 \cite{SAID} and the dotted curves to the
MAID solution 2007 \cite{MAID}.}
\end{figure*}
\begin{figure*}[pt]
\centerline{
\epsfig{file=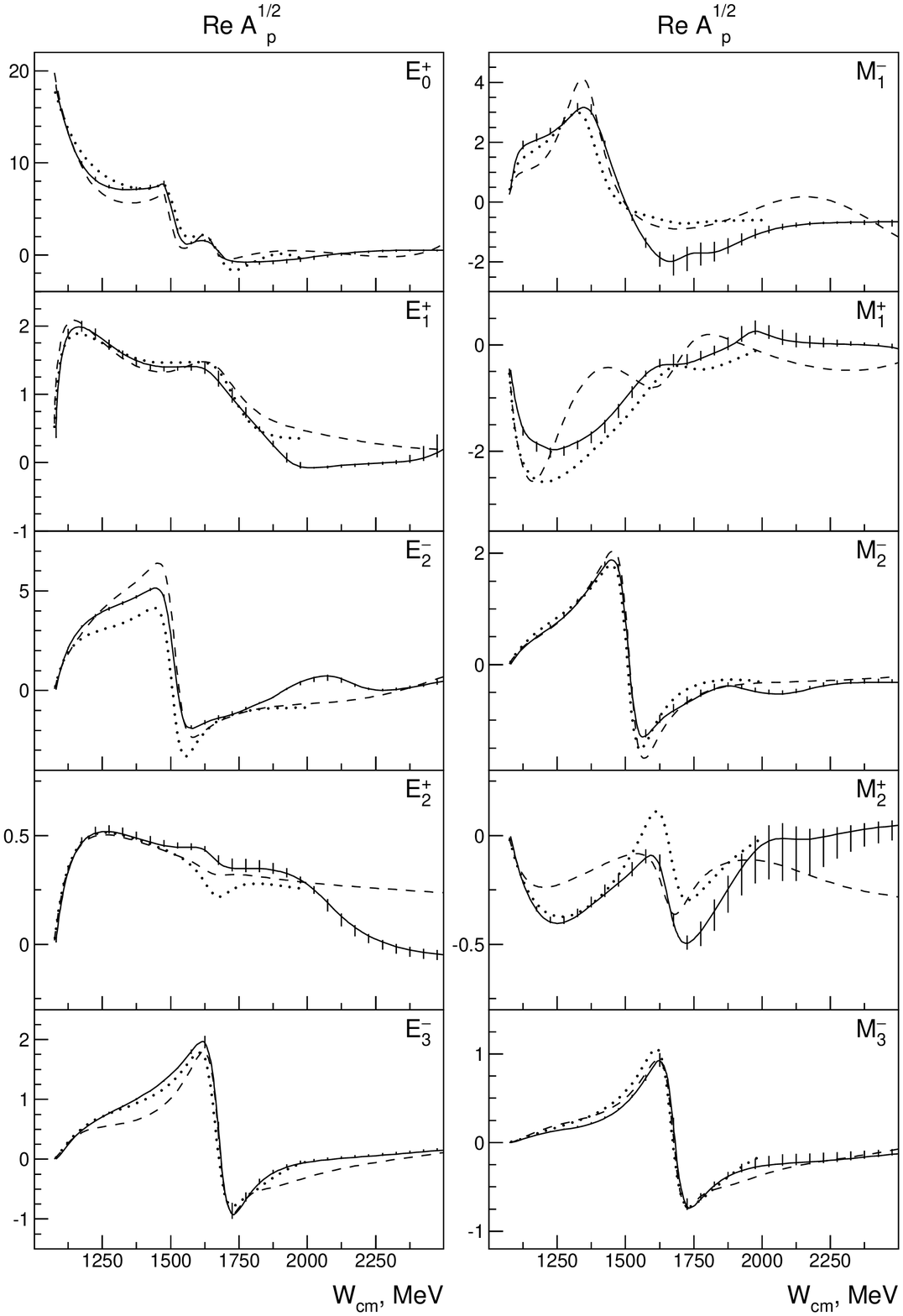,width=0.46\textwidth,height=0.45\textheight}~~
\epsfig{file=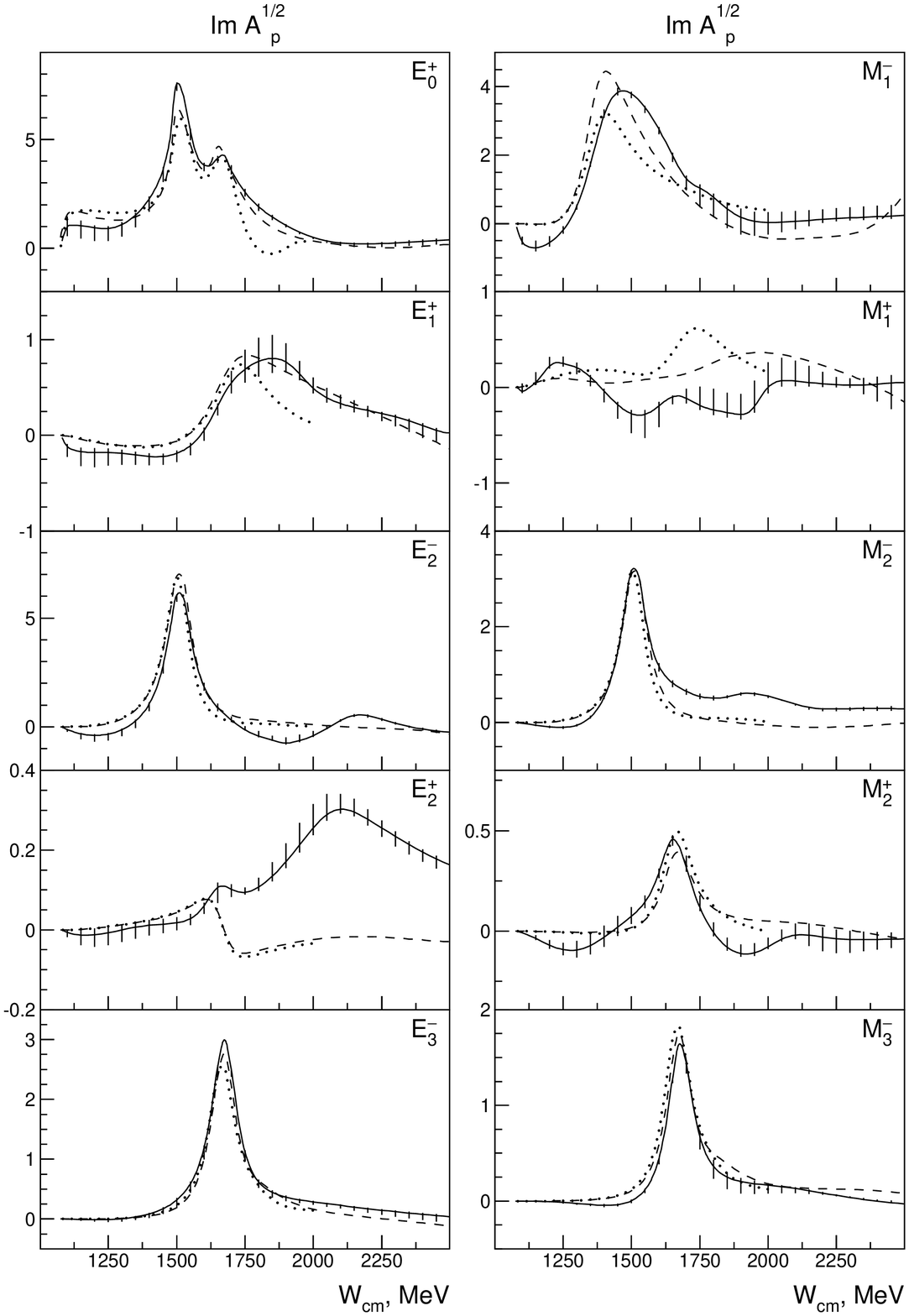,width=0.46\textwidth,height=0.45\textheight}~~
} \caption{\label{fig:pi0_mult_re} The real (two left-hand columns)
and imaginary (two right-hand columns) part of the isospin-1/2
photoproduction multipoles. The errors are systematic and cover a
large number of fits (see the text). The dashed curves correspond to
the SAID solution SP09K2700 \cite{SAID} and the dotted curves to the
MAID solution 2007 \cite{MAID}} \centerline{
\epsfig{file=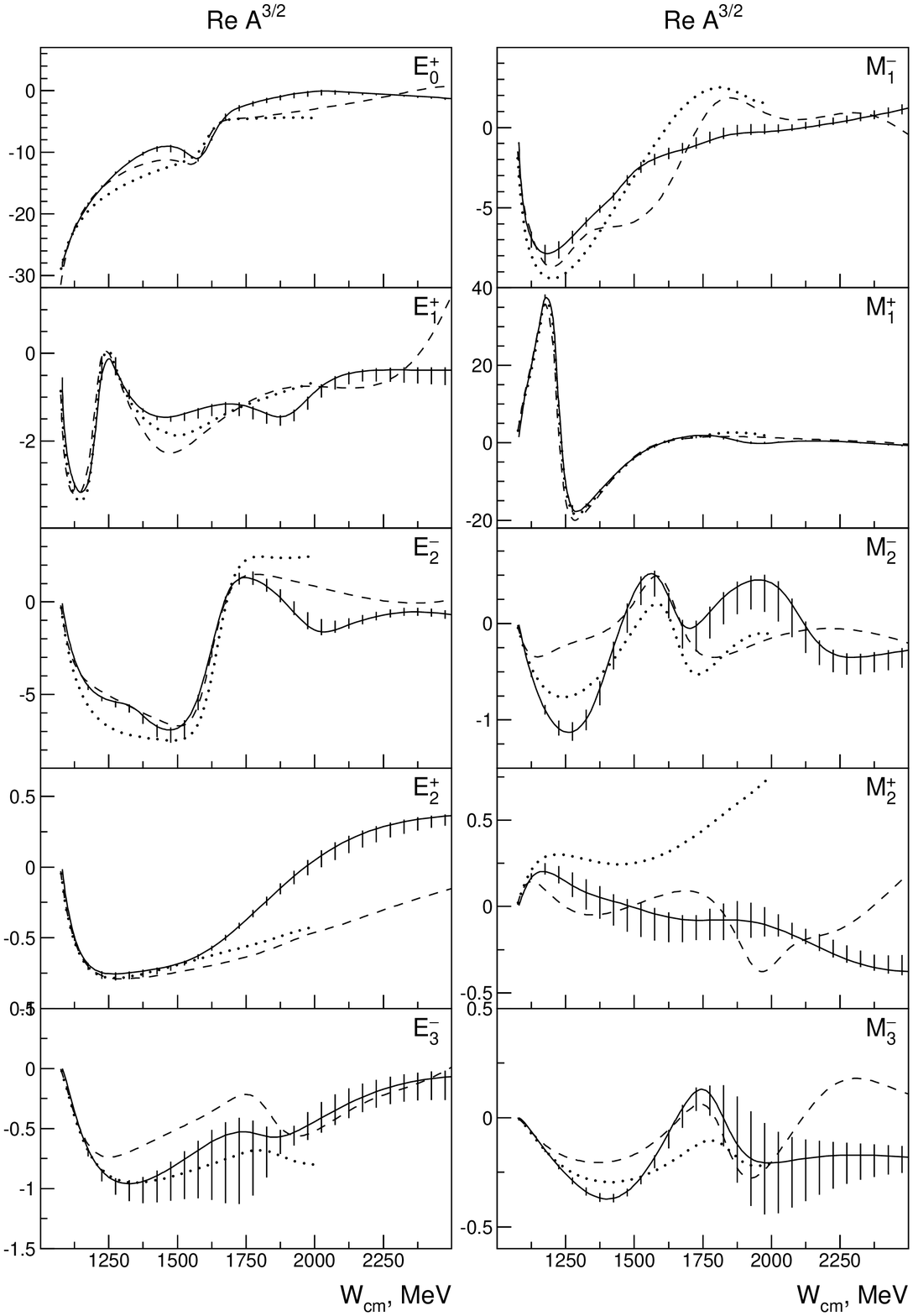,width=0.46\textwidth,height=0.45\textheight}~~
\epsfig{file=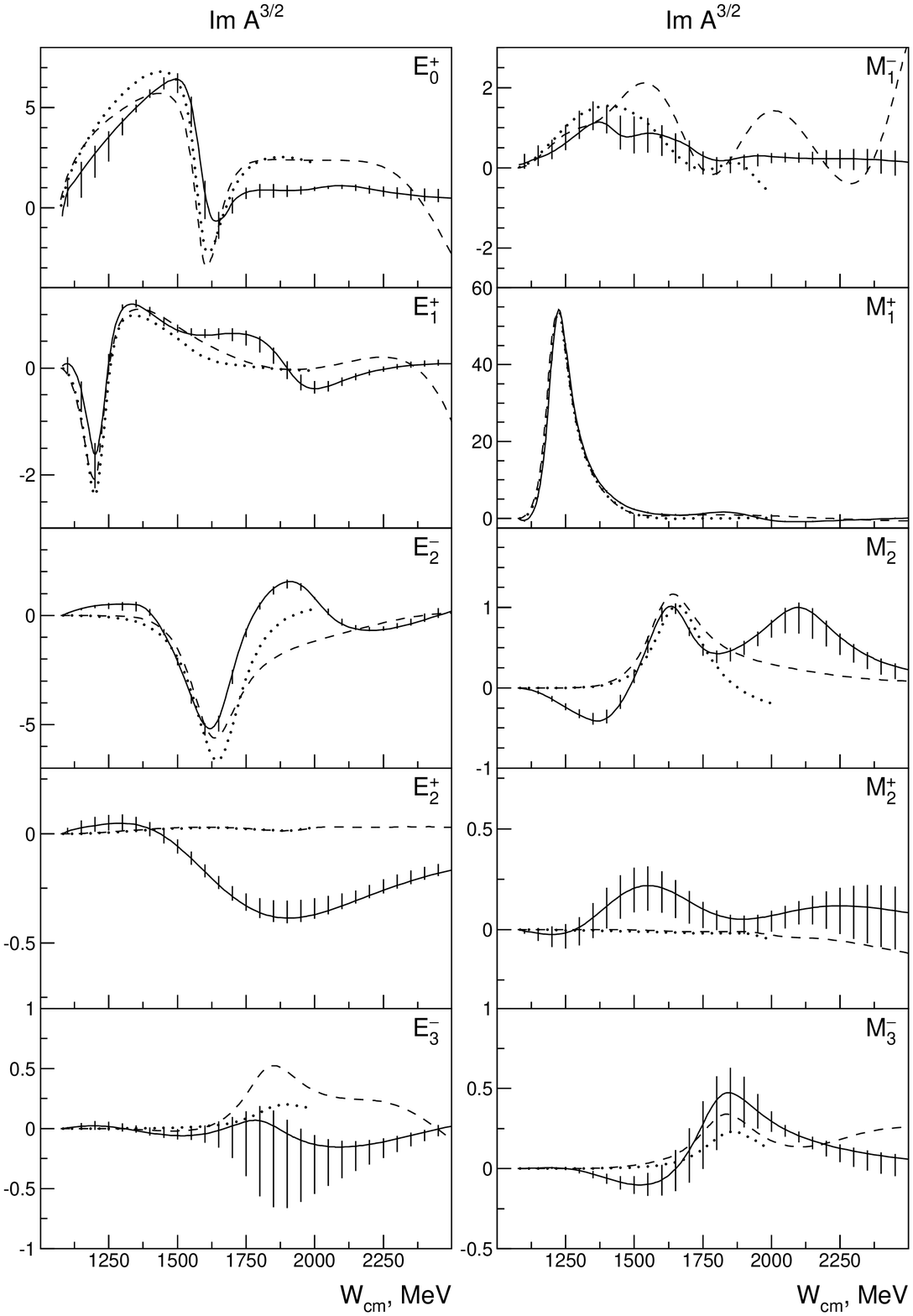,width=0.46\textwidth,height=0.45\textheight}~~
} \caption{\label{fig:pipn_mult_re} The real (two left-hand columns)
and imaginary (two right-hand columns) part of the isospin-1/2
photoproduction multipoles. The errors are systematic and cover a
large number of fits (see the text). The dashed curves correspond to
the SAID solution SP09K2700 \cite{SAID} and the dotted curves to the
MAID solution 2007 \cite{MAID}.}
\end{figure*}

\begin{table}[pb]
\caption{\label{waves}Photoproduction multipoles and partial waves.
In general, two multipoles lead to one spin-parity wave.}
\begin{center}
\renewcommand{\arraystretch}{1.2}
\begin{tabular}{cccccc}
\hline\hline \multicolumn{2}{c}{Multipoles} &
\multicolumn{2}{c}{Partial waves}&$J^P$\\\hline
$E_0^+$& - & $S_{11}$ & $S_{31}$&$1/2^-$\\
- & $M_1^-$ & $P_{11}$ & $P_{31}$&$1/2^+$\\
$E_1^+$& $M_1^+$ & $P_{13}$ & $P_{33}$&$3/2^+$\\
$E_2^-$& $M_2^-$ & $D_{13}$ & $D_{33}$&$3/2^-$\\
$E_2^+$& $M_2^+$ & $D_{15}$ & $D_{35}$&$5/2^-$\\
$E_3^-$& $M_3^-$ & $F_{15}$ & $F_{35}$&$5/2^+$\\
\hline\hline
\end{tabular}
\renewcommand{\arraystretch}{1.0}
\end{center}
\end{table}

Most amplitudes derived within the SAID, MAID, or BnGa approach
yield consistent results, at least qualitatively. The best agreement
is found for the $M_1^+$ amplitude which describes the spin flip
amplitude for the photo-induced transition from the proton to the
$\Delta$ resonance and its excitations. The $\Delta$ resonance is
fully elastic, hence the agreement in the low-mass region is not
unexpected. Even the small $E_1^+$ multipoles are not inconsistent.
Some multipoles which we discuss next show significant differences
between the different approaches. The $E_0^+$ multipole has a
similar structure in all three approaches but shows significant
differences in detail. In the BnGa solution, the electric dipole
transition $E_0^+$ exceeds the other solutions in the threshold
regions, likely due to a larger role of the subthreshold $\Lambda
K^+$ amplitude.  The differences are even larger for the $M_1^-$
multipole; this may be not unexpected in view of the notorious
difficulties with the $1/2^+$ partial wave. Surprisingly, the
multipoles for $\gamma p\to n\pi^+$ are in much better consistency.
The differences in the $E_2^-$ and $M_2^-$ can be assigned to
additional $\Delta_{3/2^-}(1940)$ and $\Delta_{3/2^-}(2260)$
resonances introduced to fit data on $\gamma p\to p\pi^0\eta$
\cite{Horn:2007pp,Horn:2008qv}. Significantly different are the
multipoles leading to $5/2^-$ states. In our fits, the $E_2^+$ and
$M_2^+$ multipoles include an additional resonance $N_{5/2^-}(2060)$
\cite{Anisovich:2005tf}. We note that the (dominant) resonance
contributions are compatible with the Watson theorem.

\subsection{Properties of contributing resonances}

A large number of resonances is identified in the fits. Some have a
strong coupling to pion photoproduction, for others, the product of
squared photocoupling constant and $N\pi$ decay branching ratio is
small and they contribute mostly to inelastic channels; their
properties will be discussed elsewhere. These latter resonances are
listed in Table~\ref{allb}. They do help to improve the fit to pion
photoproduction but their helicity amplitudes are not well defined,
and photo-couplings and decay branching ratios of these states can
be varied within large limits without significant $\chi^2$
deterioration. All these solutions were included in the error
estimation procedure.

The pole position of the states, photo-couplings and $\pi N$
branching ratios for the states contributing strongly to pion
photoproduction are given in Table~\ref{resonances}. The inclusion
of the new CLAS data rather notably stabilized the solution and
improved most of the errors. The pole positions are determined by
finding zeros of the real and the imaginary part of the denominator
in the partial wave amplitudes. Thus two lines are defined in the
complex energy plane. Their crossing points defines the pole
position. The coupling constants, including the helicity amplitudes,
are calculated as residues of the P-vector/K-matrix amplitude at the
pole position and are given together with their phases. The $\pi N$
branching ratios are calculated as squared residue-couplings,
multiplied by the phase volume taken at the Breit-Wigner resonance
mass. We note that nucleon-meson or nucleon-photon couplings are
defined at the pole position of a resonance, and are complex
numbers. In Table~\ref{resonances} we give the (complex)
photon-couplings at the pole positions; their analogues, the
helicity amplitudes $A_{1/2}$ and $A_{3/2}$ are defined for
Breit-Wigner amplitudes, not for more general formalisms. The method
how we derive Breit-Wigner parameters and helicity amplitudes is
discussed below. For the $P_{13}(1720)$ and $D_{33}(1700)$
resonances, the Breit-Wigner width is much larger than one might
expect from the pole position. In the $N\pi$ channel, the visible
width is much closer to this expectation. The effect is known from
$a_0(980)$ which has a visible width of about 50\,MeV in the
$\pi\eta$ mass distribution but a much larger width in the $K\bar K$
mass distribution the width is much larger because of the rapidly
opening $K\bar K$ phase space. In the $P_{13}(1720)$ and
$D_{33}(1700)$ case, the phase space for $N\pi\pi$ 3-body decays
grows rapidly with increasing mass.

\begin{table}[pt]
\caption{\label{allb}Baryon resonances included in the fit which
contribute little to photoproduction of pions.}
\begin{center}
\renewcommand{\arraystretch}{1.2}
\begin{tabular}{cccccc}
\hline \hline \hspace{-2mm}
\hspace{-5mm}&\hspace{-4mm}Mass\hspace{-4mm}&\hspace{-7mm}
Width&&\hspace{-4mm}Mass\hspace{-4mm}&\hspace{-7mm} Width\\\hline
\hspace{-1mm}$P_{11}(1860)$\hspace{-3mm}&\hspace{-3mm}1900$\pm
30$\hspace{-3mm}&\hspace{-3mm}300$\pm$40\hspace{-3mm}&\hspace{-2mm}
$P_{13}(1900)$\hspace{-3mm}&\hspace{-3mm}1960$\pm$30\hspace{-3mm}&\hspace{-3mm}185$\pm$40\\
\hspace{-1mm}$D_{13}(1700)$\hspace{-3mm}&\hspace{-3mm}1730$\pm$40\hspace{-3mm}&\hspace{-3mm}310$\pm$60\hspace{-3mm}&\hspace{-3mm}
$D_{13}(1875)$\hspace{-3mm}&\hspace{-3mm}1870$\pm$25\hspace{-3mm}&\hspace{-3mm}150$\pm$40\\
\hspace{-1mm}$P_{33}(1600)$\hspace{-3mm}&\hspace{-3mm}1640$\pm$40\hspace{-3mm}&\hspace{-3mm}480$\pm$1{\scriptsize
00}\hspace{-3mm}&\hspace{-3mm}
$P_{33}(1920)$\hspace{-3mm}&\hspace{-3mm}1950$\pm$40\hspace{-3mm}&\hspace{-3mm}330$\pm$50\\
\hspace{-1mm}$F_{15}(2000)$\hspace{-3mm}&\hspace{-3mm}1910$\pm$50\hspace{-3mm}&\hspace{-3mm}360$\pm$80\hspace{-3mm}&\hspace{-3mm}
$D_{15}(2070)$\hspace{-3mm}&\hspace{-3mm}2065$\pm$25\hspace{-3mm}&\hspace{-3mm}340$\pm$40\\
\hspace{-1mm}$D_{33}(1940)$\hspace{-3mm}&\hspace{-3mm}1995$\pm$40\hspace{-3mm}&\hspace{-3mm}360$\pm$50\hspace{-3mm}&\hspace{-3mm}
$D_{13}(2170)$\hspace{-3mm}&\hspace{-3mm}2160$\pm$35\hspace{-3mm}&\hspace{-3mm}370$\pm$50\\
\hspace{-1mm}$D_{33}(2360)$\hspace{-3mm}&\hspace{-3mm}2360$\pm$50\hspace{-3mm}&\hspace{-3mm}480$\pm$80&
\hspace{-3mm}
$S_{31}(1900)$\hspace{-3mm}&\hspace{-3mm}1955$\pm$30\hspace{-3mm}&\hspace{-3mm}335$\pm$40\\
 \hline\hline
\end{tabular}
\renewcommand{\arraystretch}{1.0}
\end{center}
\end{table}

\begin{table}[pt]
\caption{\label{resonances} Pole position (in MeV), photo-couplings
calculated as residues in the pole (in GeV$^{-1/2}10^{3}$, phases in
degrees) and corresponding Breit-Wigner parameters for states
contributing strongly to pion photoproduction (the branching ratios are
in percents). The PDG values are
given in parentheses. For $F_{35}(1905)$ two solutions are given.}
\begin{footnotesize}
\renewcommand{\arraystretch}{1.00}
\bc
\begin{tabular}{lcc}
\hline\hline
State                   &   $S_{11}(1535)$                 &    $S_{11}(1650)$
\\\hline
Re(pole)                & $1510\!\pm\!25$ ($1510\!\pm\!20$)& $1670\!\pm\!35$ ($1655\!\pm\!15$)\\
-2Im(pole)              & $140\!\pm\!30$ ($170\!\pm\!80$)  & $170\!\pm\!40$ ($165\!\pm\!15$)   \\
$A^{1/2}(\gamma p)$     & $90\!\pm\!25\,/0^o\!\pm\!45^o$   & $65\!\pm\!30$\,/$28^o\pm 15^o$    \\
$M_{BW}$                & $1535\!\pm\!20$ ($1535\!\pm\!10$)& $1680\!\pm\!40$ ($1658\!\pm\!12$) \\
$\Gamma_{BW}$           & $170\!\pm\!35$ ($150\!\pm\!25$) &  $170\!\pm\!45$ ($165\!\pm\!20$)  \\
$\Gamma_{\pi N}/\Gamma$ & $35\pm 15$ ($45\pm 10$)          & $50\pm
25$ ($78\pm 18$)    \\ \hline State                   &
$P_{11}(1440)$                 &  $S_{31}(1620)$
\\\hline
Re(pole)                & $1370\!\pm\!4$ ($1365\!\pm\!15$) &$1596\!\pm\!7$ ($1600\!\pm\!10$) \\
-2Im(pole)              & $193\!\pm\!7$ ($~190\!\pm\!30$)  &~$130\!\pm\!10$ ($~118\!\pm\!3$)  \\
$A^{1/2}(\gamma p)$     & -$48\!\pm\!12$\,/-$58^o\!\pm\!20^o$ &$62\!\pm\!10$\,/-$0^o\!\pm\!20^o$     \\
$M_{BW}$                & $1440\!\pm\!12$ ($1445\!\pm\!25$)& $1625\!\pm\!10$ ($1630\!\pm\!30$)   \\
$\Gamma_{BW}$           & $335\!\pm\!50$($325\!\pm\!125$)&  $148\!\pm\!15$ ($143\!\pm\!8$)   \\
$\Gamma_{\pi N}/\Gamma$ & $60\pm 6$ ($65\pm 15$)           & $23\pm 5$ ($25\pm 5$)           \\
\hline State                   &  $P_{11}(1710)$                  &
$P_{33}(1232)$                  \\\hline
Re(pole)                &$1708\!\pm\!18$ ($1720\!\pm\!50$) &$1211\!\pm\!1$ ($1210\!\pm\!1$) \\
-2Im(pole)              &$200\!\pm\!20$ ($230\!\pm\!150$)  &$100\!\pm\!2$ ($~100\!\pm\!2$)   \\
$A^{1/2}(\gamma p)$     &$24\!\pm\!8$\,/-$20^o\!\pm\!60^0$ &-$136\!\pm\!5$\,/-$17^o\!\pm\!5^o$  \\
$A^{3/2}(\gamma p)$     &     ~                            &-$267\!\pm\!8$\,/-$3^o\!\pm\!3^o$  \\
$M_{BW}$                &$1725\!\pm\!25$ ($1710\!\pm\!30$) &$1230\!\pm\!2$ ($1232\!\pm\!1$)   \\
$\Gamma_{BW}$           &$200\!\pm\!35$ ($150\!\pm\!100$) &$112\!\pm\!4$ ($118\!\pm\!2$)   \\
$\Gamma_{\pi N}/\Gamma$ &$12\pm 6$ ($15\pm 5$)            & $100$ ($100$)                    \\
\hline State                   &  $P_{13}(1720)$                  &
$D_{33}(1700)$                  \\\hline
Re(pole)                &$1660\!\pm\!35$ ($1675\!\pm\!15$) &$1650\!\pm\!30$ ($1650\!\pm\!30$)\\
-2Im(pole)              &$360\!\pm\!80$ ($190\!\pm\!85$)   &$275\!\pm\!35$ ($200\!\pm\!40$)  \\
$A^{1/2}(\gamma p)$     &$140\!\pm\!50$\,/-$35^o\!\pm\!25^o$&$160\!\pm\!45$\,/$35^o\!\pm\!12^o$\\
$A^{3/2}(\gamma p)$     &$110\!\pm\!50$\,/$10^o\!\pm\!35^o$&$165\!\pm\!40$\,/$40^o\!\pm\!18^o$\\
$M_{BW}$                & $1770\!\pm\!100$ ($1725\!\pm\!25$) &$1780\!\pm\!40$ ($1710\!\pm\!40$)  \\
$\Gamma_{BW}$           &$650\!\pm\!120$ ($225\!\pm\!75$)  & $580\!\pm\!120$ ($300\!\pm\!100$)   \\
$\Gamma_{\pi N}/\Gamma$ & $14\pm 5$ ($15\pm 5$)           &$16\pm 7$ ($15\pm 5$)            \\
\hline State                   &  $D_{13}(1520)$ & $F_{35}(1905)$ (sol.1)\\
\hline
Re(pole)                &$1512\!\pm\!3$ ($1510\!\pm\!5$)  &$1800\!\pm\!15$ ($1830\!\pm\!5$)  \\
-2Im(pole)              &$110\!\pm\!6$ ($112\!\pm\!7$)   &$300\!\pm\!20$ ($282\!\pm\!18$)  \\
$A^{1/2}(\gamma p)$     &-$30\!\pm\!6$\,/$15^o\!\pm\!10^o$&$28\!\pm\!10$\,/-$35^o\!\pm\!15^o$ \\
$A^{3/2}(\gamma p)$     &$130\!\pm\!6$\,/$6^o\!\pm\!5^o$  &-$42\!\pm\!12$\,/-$25^o\!\pm\!15^o$ \\
$M_{BW}$                &$1524\!\pm\!4$ ($1520\!\pm\!5$) & $1890\!\pm\!25$ ($1890\!\pm\!25$)    \\
$\Gamma_{BW}$           &$117\!\pm\!6$ ($112\!\pm\!13$) & $335\!\pm\!30$ ($335\!\pm\!65$)    \\
$\Gamma_{\pi N}/\Gamma$ &    $57\pm 5$ ($60\pm 5$)        & $12\pm 3$ ($12\pm 3$)           \\
\hline State                   &  $D_{15}(1675)$                 &
$F_{35}(1905)$ (sol.2)               \\\hline
Re(pole)                &$1650\!\pm\!5$ ($1660\!\pm\!5$)  &$1805\!\pm\!15$ ($1830\!\pm\!5$) \\
-2Im(pole)              &$143\!\pm\!7$ ($~138\!\pm\!12$)  &$310\!\pm\!20$ ($282\!\pm\!18$)  \\
$A^{1/2}(\gamma p)$     &$20\!\pm\!4$\,/-$6^o\!\pm\!6^o$ &$47\!\pm\!10$\,/-$30^o\!\pm\!12^o$ \\
$A^{3/2}(\gamma p)$     &$24\!\pm\!8$\,/-$6^o\!\pm\!6^o$ & $0\!\pm\!3$ \\
$M_{BW}$                &$1678\!\pm\!5$ ($1675\!\pm\!5$) & $1850\!\pm\!20$ ($1890\!\pm\!25$)  \\
$\Gamma_{BW}$           &$177\!\pm\!15$ ($148\!\pm\!18$) & $345\!\pm\!30$ ($335\!\pm\!65$)   \\
$\Gamma_{\pi N}/\Gamma$ & $37\pm 5$ ($40\pm 5$)           & $12\pm 3$ ($12\pm 3$)             \\
\hline State                   &$F_{15}(1680)$                   &
$F_{37}(1950)$                \\\hline
Re(pole)                &$1672\!\pm\!4$ ($1673\!\pm\!8$)  & $1882\!\pm\!8$ ($1880\!\pm\!10$) \\
-2Im(pole)              &$114\!\pm\!12$ ($133\!\pm\!12$)  & $262\!\pm\!12$ ($240\!\pm\!20$)  \\
$A^{1/2}(\gamma p)$     &-$12\!\pm\!6$\,/-$45^o\!\pm\!30^o$&-$81\!\pm\!8$\,/-$15^o\!\pm\!12^o$ \\
$A^{3/2}(\gamma p)$     &$130\!\pm\!8$\,/$0^o\!\pm\!10^o$ &-$93\!\pm\!8$\,/-$15^o\!\pm\!15^o$  \\
$M_{BW}$                &$1685\!\pm\!5$ ($1685\!\pm\!5$)  & $1928\!\pm\!8$ ($1933\!\pm\!18$)   \\
$\Gamma_{BW}$           &$117\!\pm\!12$ ($130\!\pm\!10$)  & $290\!\pm\!14$($285\!\pm\!50$)    \\
$\Gamma_{\pi N}/\Gamma$ & $66\pm 8$ ($68\pm 3$)           &  $44\pm 8$ ($40\pm 5$)            \\
\hline\hline
\end{tabular}
\ec
\end{footnotesize}
\renewcommand{\arraystretch}{1.0}
\end{table}
Within the quoted errors, the results from the new solution are
mostly compatible with those published in \cite{Thoma:2007bm}. The
largest changes to our previous solution are observed for the
photo-couplings of the $S_{31}(1620)$ resonance (which was (130$\pm$
50)\,GeV$^{-1/2}$$\times$$10^{3}$ in \cite{Thoma:2007bm}), and for
the small helicity component $A_{1/2}$ of the $D_{13}(1520)$ (which
was (7.0$\pm$1.5)\,GeV$^{-1/2}$$\times$$10^{3}$). The changes are
largely due to the inclusion of additional polarization data for
$\gamma p\to \pi^0 p$ and $\gamma p \to \pi^+ n$.

The new data also require $P_{11}(1710)$. In
\cite{Sarantsev:2007bk}, this resonance improved the description of
the data slightly but we were not forced to introduce it. In the
present fit, there are three resonances above the nucleon in the
$P_{11}$ wave: the Roper resonance $P_{11}(1440)$, the
$P_{11}(1710)$, and the newly proposed $P_{11}(1860)$.

We found two minima for the photo-couplings of the $F_{35}(1905)$
state. Both solutions are given in the Table~\ref{resonances}
($5^{\rm th}$ and $6^{\rm th}$ row, $2^{\rm nd}$ column). The first
solution corresponds well to the PDG average values, while the
second solution has an almost vanishing helicity-3/2 coupling. Both
solutions reproduce single meson photoproduction data with the same
quality, however the second solution provides a better likelihood
for the $\gamma p\to \pi^0\eta p$ reaction. The analysis of the high
energy data on the two-pion photoproduction will help to define
which solution is the physical one. Also the forthcoming double
polarization data will provide new and important constraints on the
amplitudes.

The pole structure of two resonances is found to be ambiguous. The
pole of the $S_{11}(1650)$ resonance is located between the $\Lambda
K$ and $\Sigma K$ thresholds. In one set of acceptable solutions we
found large couplings of this state to these channels and a
complicated pole structure of two or even more poles close to the
two thresholds. We included such pole positions and corresponding
residues in the errors given in Table~\ref{resonances}. The
forthcoming analysis of the data on the $\pi p\to K\Lambda$  should
help to define the $S_{11}$ pole structure more accurately, however
new data on $\pi N\to K\Lambda$ and $\pi N\to K\Sigma$ would be
extremely valuable. The fit of two-pion photoproduction data
suggests a substantial coupling of $P_{13}(1720)$ to the
$D_{13}(1520)\pi$ channel. Thus we observe here a double pole
structure near the $D_{13}(1520)\pi$ threshold which results in
rather large errors and in difficulties to identify the
corresponding Breit-Wigner parameters. We believe that the
forthcoming polarization data on two-pion photoproduction will
improve significantly the accuracy in the definition of
$P_{13}(1720)$ pole structure. A more precise definition of the
$P_{13}(1720)$ pole with its $N\pi\pi$ couplings
\begin{table*}[pt]
\caption{\label{helicities}Helicity amplitudes $A_{1/2}$ and
$A_{3/2}$ for $N^\ast$ and $\Delta^\ast$ from this work, from SAID08
~\protect\cite{SAID}, from MAID07~\protect\cite{MAID}, from the
Gie\ss en model \protect\cite{Penner:2002md,Shklyar:2006xw} and
estimates from Ref.~\protect\cite{Amsler:2008zzb}.} \vspace{2mm}
\renewcommand{\arraystretch}{1.2}
\bc\begin{tabular}{rrrrrrrrrrr} \hline\hline   Resonance        &
\multicolumn{5}{c}{$A_{1/2}$ (GeV$^{-1/2}$$\times$$10^{3}$)} &
\multicolumn{5}{c}{$A_{3/2}$ (GeV$^{-1/2}$$\times$$10^{3}$)} \\
  & BnGa09 &\hspace{-2mm} FA08\hspace{2mm} &MAID07\hspace{-2mm}&\hspace{-2mm}Gie\ss en\hspace{-4mm}&\hspace{-2mm}PDG\hspace{2mm}&\hspace{-4mm}BnGa09\hspace{4mm}& \hspace{-3mm} FA08 \hspace{3mm} &\hspace{-2mm}MAID07\hspace{-2mm}&\hspace{-2mm}Gie\ss en\hspace{-2mm}&\hspace{-2mm} PDG \\
\hline\hline
$S_{11}(1535)$      &$90\!\pm\!15$ &    100.9$\pm$3.0 &\hspace{-4mm}    66\hspace{4mm} &\hspace{-2mm}95 \hspace{2mm} &   90$\pm$30 && && &\\
$S_{11}(1650)$      &$60\!\pm\!20$ &      9.0$\pm$9.1 &\hspace{-4mm}    33\hspace{4mm} &\hspace{-2mm}57 \hspace{2mm} &   53$\pm$16 && && &\\
$P_{11}(1440)$      &-$52\!\pm\!10$&  $-$56.4$\pm$1.7 &\hspace{-4mm} $-$61\hspace{4mm} &\hspace{-2mm}-84 \hspace{2mm} & $-$65$\pm$4  && && &\\
$P_{11}(1710)$      &$25\!\pm\!10$&&                          &\hspace{-2mm}-50 \hspace{2mm} & $9\!\pm\!22$   && && &\\
$P_{13}(1720)$      &$130\!\pm\!50$&     90.5$\pm$3.3 &\hspace{-4mm}    73\hspace{4mm} &\hspace{-2mm}-65 \hspace{2mm} &    18$\pm$30 &$100\!\pm\!50$& $-$36.0$\pm$3.9 & \hspace{-2mm} $-$11 \hspace{2mm} & \hspace{-2mm} 35 \hspace{2mm} & $-$19$\pm$20 \\
$D_{13}(1520)$      &-$32\!\pm\!6$ &    $-$26$\pm$1.5 &\hspace{-4mm} $-$27\hspace{4mm} &\hspace{-2mm}-15 \hspace{2mm} & $-$24$\pm$9  &$138\!\pm\!8$&   141.2$\pm$1.7 &  \hspace{-2mm}  161 \hspace{2mm}  & \hspace{-2mm} 146 \hspace{2mm} &   166$\pm$5 \\
$D_{15}(1675)$      &$21\!\pm\!4$  &     14.9$\pm$2.1 &\hspace{-4mm}    15\hspace{4mm} &\hspace{-3mm}9 \hspace{3mm} &    19$\pm$8  &$24\!\pm\!8$&    18.4$\pm$2.1 & \hspace{-2mm} 22 \hspace{2mm} & \hspace{-2mm} 21 \hspace{2mm} &  15$\pm$9 \\
$F_{15}(1680)$      &-$12\!\pm\!6$ &  $-$17.6$\pm$1.5 &\hspace{-4mm} $-$25\hspace{4mm} &\hspace{-3mm}3  \hspace{3mm} &  $-$15$\pm$6  &$136\!\pm\!12$&   134.2$\pm$1.6 & \hspace{-2mm} 134 \hspace{2mm} & \hspace{-2mm} 116 \hspace{2mm} &   133$\pm$12 \\
\hline
$S_{31}(1620)$ &$63\!\pm\!12$&     47.2$\pm$2.3 &\hspace{-4mm}    66\hspace{4mm}       &\hspace{-2mm} -50\hspace{2mm} &    27$\pm$11 && & &\\
$P_{33}(1232)$ &-$136\!\pm\!5$& $-$139.6$\pm$1.8 &\hspace{-4mm} $-$140\hspace{4mm}     &\hspace{-2mm} -128\hspace{2mm} & $-$135$\pm$6 &-$267\!\pm\!8$ & $-$258.9$\pm$2.3& \hspace{-2mm} $-$265 \hspace{2mm} & \hspace{-2mm} -247 \hspace{2mm} & $-$250$\pm$8 \\
$D_{33}(1700)$ &$160\!\pm\!45$&    118.3$\pm$3.3 &\hspace{-4mm}   226\hspace{4mm}      &\hspace{-2mm} 96\hspace{2mm} &    104$\pm$15&$160\!\pm\!40$&    110.0$\pm$3.5& \hspace{-2mm} 210 \hspace{2mm} & \hspace{-2mm} 154 \hspace{2mm} &     85$\pm$22 \\
$F_{35}(1905)$ &$28\!\pm\!12$&     11.4$\pm$8.0 &\hspace{-4mm}    18\hspace{4mm}       &\hspace{-2mm} \hspace{2mm} &     26$\pm$11&-$42\!\pm\!15$&  $-$51.0$\pm$8.0& \hspace{-2mm} $-$28 \hspace{2mm} &&  $-$45$\pm$20 \\
or: &($48\!\pm\!12$)&   &     &&  &($0\!\pm\!3$)&  & &   \\
$F_{37}(1950)$ &-$83\!\pm\!8$&  $-$71.5$\pm$1.8 &\hspace{-4mm} $-$94\hspace{4mm}       &\hspace{-2mm} \hspace{2mm} &  $-$76$\pm$12&-$92\!\pm\!8$&  -$96\!\pm\!8$ & \hspace{-2mm} $-$121 \hspace{2mm} &&  $-$97$\pm$10 \\
\hline\hline
\end{tabular}\ec
\end{table*}\noindent
may also help to resolve the discrepancies in the determination of
its $A_{3/2}$ helicity amplitude when different analyses are
compared.

The Review of Particle Properties lists Breit-Wigner parameters and
real helicity amplitudes. In the K-matrix approach, a photo-produced
resonance is described by a P-vector (eq.~\ref{Pvect}). Even if the
photo-coupling constant $g_{\gamma N}^{(\alpha)}$ is real, the
photo-coupling at the resonance position, calculated as residuum of
the amplitude $P_b$ at the pole, will in general be complex. To
allow for a comparison with other determinations, we define helicity
amplitudes by the following procedure: a Breit-Wigner amplitude is
constructed with an adjustable mass and a width which is
parameterized as a sum of all partial widths, $\sum \rho_i g_i^2$.
The total widths is scaled with one parameter. This scaling
parameter as well as the Breit-Wigner mass are adjusted to reproduce
the pole position of the P-vector amplitude. The Breit-Wigner
helicity amplitudes $A_{1/2}$ and $A_{3/2}$ are defined by the
condition that the residues of the Breit-Wigner photoproduction
amplitude reproduce the magnitude of the original residues of the
P-vector/K-matrix amplitude.

In Table \ref{helicities} we compare our results on $A_{1/2}$ and
$A_{3/2}$ for $N^\ast$ and $\Delta^\ast$ with previous
determinations of these quantities. These real helicity amplitudes
are given in Table \ref{helicities} and compared to values obtained
by SAID, MAID, and the Gie\ss en model, and to the values listed by
the PDG \cite{Amsler:2008zzb}. First we notice that our errors are
much larger than those given by FA08, MAID and Gie\ss en do not give
any errors. We believe that the FA08 systematic errors are
underestimated: the impact of variations in the couplings to
inelastic channels can hardly be tested using only reactions with
$N\pi$ in the final state. The errors we quote are not statistical
errors; those are small. Our errors are derived from a large number
of fits changing the number of resonances, switching on and off
couplings to inelastic channels, using different start values for
the fits.

For most resonances, reasonable consistency between the different
analyses is found. In particular the helicity amplitudes for
photoproduction of the Roper resonance from SAID, MAID, and BnGa are
fully consistent (the Gie\ss en result is a bit higher) even though
mass, width, and $N\pi$ decay branching fractions differ somewhat.
BnGa and SAID, e.g., find,
respectively,\\[-3ex]

\bc
\renewcommand{\arraystretch}{1.2}\begin{tabular}{cccc}
                  & M (MeV)& $\Gamma$ (MeV)&$\Gamma_{N\pi}/\Gamma_{\rm
                  tot}$\\\hline
 BnGa & $1440\pm 12$&$335\pm 50$&$0.60\pm 0.06$\\
FA08& 1485& 284  & 0.79\\\hline
\end{tabular}\ec
\noindent In our analysis, the Roper resonance is fully constrained:
from three of the four reactions, $\pi N$ elastic scattering,
$\gamma p\to N\pi$, $\pi^-p\to p\pi^0\pi^0$, and $\gamma p\to
p\pi^0\pi^0$, the amplitude for the forth reaction can be predicted.
Hence we are particularly confident that these results are correct.

We comment briefly on further differences. The PDG result for the
$A_{1/2}$ amplitude of (53$\pm$16)\,GeV$^{-1/2}$$\times$$10^{3}$ for
producing $S_{11}(1650)$ was driven by the 1995 VPI result
(69$\pm$5)\,GeV$^{-1/2}$$\times$$10^{3}$ \cite{Arndt:1995ak} and by
the small value (22$\pm$7)\,GeV$^{-1/2}$$\times$$10^{3}$ obtained in
\cite{Dugger:2007bt}. The most recent FA08 analysis gives
(9.0$\pm$9.1)\,GeV$^{-1/2}$$\times$$10^{3}$, a value which is much
smaller and which is not confirmed here; we find
(60$\pm$20)\,GeV$^{-1/2}$ $\times$$10^{3}$, in close agreement with
the Gie\ss en result. Part of the discrepancy with FA08 is certainly
due to the $S_{11}(1650)$ branching ratio to the $\pi N$ channel; in
FA08 this is fixed to be 100\% while we find (50$\pm$25)\%. Of
course, photoproduction defines only the product of the helicity and
$\pi N$ couplings.

Possibly related are the differences in the helicity amplitudes for
$P_{13}(1720)$. Our value for $A_{1/2}$ is compatible with the new
FA08 analysis and in conflict with the value quoted by the PDG.
Incompatible with all other determinations  - even in the sign - is
our value for the $A_{3/2}$ helicity amplitude for $P_{13}(1720)$
production. Also the Gie\ss en results are at variance with the
other determinations. Clearly, more data are required to resolve
this discrepancy; the results from double polarization experiments
carried out at present in different laboratories will very likely be
decisive.

There is the possibility, that the discrepancies in the properties
of $P_{13}(1720)$ have a physical origin. In \cite{Ripani:2002ss} it
was found that data on the reaction $e p \to e' p \pi^+\pi^-$ could
be described only when resonance parameters were drastically changed
with respect to published results, or when a new resonance in the
$P_{13}$ wave was introduced. Apparently, the $P_{13}(1720)$
properties are different in $N\pi$ and in $N\pi\pi$; this might be a
hint for the presence of a close-by state in the same partial wave.

\section{Summary}
We have presented results from a partial wave analysis on a large
variety of different reactions, from $\pi N$ elastic scattering to
photoproduction of multibody final states. The main emphasis of this
paper was devoted to a determination of the electric and magnetic
multipoles leading to the production of neutral or charged pions in
photo-induced reactions off protons. The multipoles are mostly
consistent with previous analyses but a few significant
discrepancies call for clarifications. The analysis provides masses,
widths, and helicity amplitudes for several known resonances. Masses
and widths and the $\pi N$ partial decay widths of all resonances
agree very well with established values. Only the photocoupling of
the $P_{13}(1720)$ resonance differs remarkably from PDG  and
from the values found in a recent analysis of the CLAS
collaboration. This discrepancy may be a further hint for the
conjecture \cite{Ripani:2002ss} that the $P_{13}(1720)$ resonance
may have a more complicated structure than usually assumed.

\subsection*{Acknowledgements}
We would like to thank the members of SFB/TR16 for continuous
encouragement. We acknowledge financial support from the Deutsche
Forschungsgemeinschaft (DFG) within the SFB/TR16 and from the
Forschungszentrum J\"ulich within the FFE program. The collaboration
with St. Petersburg received funds from DFG and the Russian
Foundation for Basic Research.

\section*{Appendix A: The structure of the fermion propagator}
We consider scattering of two particles with momenta $k_1$ and $k_2$
in the initial and $q_1$ and $q_2$ in the final state. There are
three independent momenta. It is convenient to choose the total
four-momentum of the system $P=k_1+k_2=q_1+q_2$ and two relative
momenta $k^\perp_\mu$ and $q^\perp_\mu$ which are orthogonal to the
total momentum:
\be
k^\perp_\mu&=&\frac 12 (k_1-k_2)_\nu g_{\mu\nu}^\perp,\qquad
q^\perp_\mu=\frac 12 (q_1-q_2)_\nu g_{\mu\nu}^\perp, \nn
&&g^\perp_{\mu\nu}=\left (g^\perp_{\mu\nu}-\frac{P_\mu
P_\nu}{P^2}\right ).
\ee

The tensor $F^{\mu_1\ldots\mu_n}_{\nu_1\ldots\nu_n}$ depends only on
the total momentum $P$ ($s=P^2$) and describes the tensor structure
of the partial wave. It can be calculated as a product of two
polarization tensors $\Psi^{\alpha}_{\nu_1\ldots\nu_n}$ summed over
possible polarizations:
\be
F^{\mu_1\ldots\mu_n}_{\nu_1\ldots\nu_n}=\sum\limits_\alpha
\Psi^{\alpha *}_{\mu_1\ldots\mu_n}
\Psi^{\alpha}_{\nu_1\ldots\nu_n}\,.
\ee
For every set of indices, $F^{\mu_1\ldots\mu_n}_{\nu_1\ldots\nu_n}$
satisfies the properties of the polarization tensor: it is
symmetrical over permutation of any two indices, traceless and for
$n>0$ is orthogonal to the total momentum of the system. It usually
normalized by the condition:
\be
F^{\mu_1\ldots\mu_n}_{\nu_1\ldots
\nu_n}F^{\nu_1\ldots\nu_n}_{\xi_1\ldots\xi_n}=
(-1)^nF^{\mu_1\ldots\mu_n}_{\xi_1\ldots\xi_n}.
\ee
and is often called projection operator: its convolution with
another tensor by one set of indices results in a tensor which obeys
the symmetry properties of the corresponding partial wave.

In the case of a fermionic system,
$F^{\mu_1\ldots\mu_n}_{\nu_1\ldots\nu_n}$ can be written in the form
\be
F^{\mu_1\ldots\mu_n}_{\nu_1\ldots\nu_n}\!=\!(-1)^n
\frac{\sqrt{s}\!+\!\hat P}{2\sqrt{s}}
O^{\mu_1\ldots\mu_n}_{\xi_1\ldots \xi_n}
T^{\xi_1\ldots\xi_n}_{\beta_1\ldots \beta_n} O^{\beta_1\ldots
\beta_n}_{\nu_1\ldots\nu_n}\,.
\label{fp}
\ee
Here, $(\sqrt s+\hat P)$ corresponds to the numerator of a
propagator describing a particle with $J=1/2$ and $n\!=\!J\!-\!1/2$
($\sqrt s\!=\!M$ for the stable particle). We define
\bea
T^{\xi_1\ldots\xi_n}_{\beta_1\ldots \beta_n}&=& \frac{n+1}{2n\!+\!1}
\big( g_{\xi_1\beta_1}\!-\! \frac{n}{n\!+\!1}\sigma_{\xi_1\beta_1}
\big) \prod\limits_{i=2}^{n}g_{\xi_i\beta_i},
\\ \nn
\sigma_{\alpha_i\alpha_j}&=&\frac 12
(\gamma_{\alpha_i}\gamma_{\alpha_j}-
\gamma_{\alpha_j}\gamma_{\alpha_i}).
\label{t1}
\eea
We introduced the factor $1/(2\sqrt s)$ in the propagator which
removes the divergency of this function at large energies. For the
stable particle it means that bispinors are normalized as follows:
\bea
 \bar u(k_N) u(k_N)\!=\!1\;,\;\;
\sum\limits_{polarizations}\!\!\!\!\!\! u(k_N)\bar u(k_N)
\!=\!\frac{m\!+\!\hat k_N}{2m}\;.
\label{bisp_norm}
\eea
Here and below, $\hat k\equiv\gamma_\mu k_\mu$.

The boson projection operator $O^{\mu_1\ldots\mu_n}_{\nu_1\ldots
\nu_n}$ has the following properties:
\bea
&&P_{\mu_i}O^{\mu_1\ldots\mu_n}_{\nu_1\ldots \nu_n}
=P_{\nu_j}O^{\mu_1\ldots\mu_n}_{\nu_1\ldots \nu_n}=0\;, \nn
&&g_{\mu_i\mu_j}O^{\mu_1\ldots\mu_n}_{\nu_1\ldots \nu_n}
=g_{\nu_i\nu_j}O^{\mu_1\ldots\mu_n}_{\nu_1\ldots \nu_n}=0\;, \nn
&&O^{\mu_1\ldots\mu_n}_{\alpha_1\ldots \alpha_n} O^{\alpha_1\ldots
\alpha_n}_{\nu_1\ldots \nu_n}= (-1)^n
O^{\mu_1\ldots\mu_n}_{\nu_1\ldots \nu_n} \;.
\label{O_proper}
\eea
For the lowest states,
\be
O\!&=&\! 1\ ,\qquad O^\mu_\nu\!=\!g_{\mu\nu}^\perp\ , \nn
O^{\mu_1\mu_2}_{\nu_1\nu_2}\!&=&\! \frac 12 \left (
g_{\mu_1\nu_1}^\perp  g_{\mu_2\nu_2}^\perp \!+\!
g_{\mu_1\nu_2}^\perp  g_{\mu_2\nu_1}^\perp  \!- \!\frac 23
g_{\mu_1\mu_2}^\perp  g_{\nu_1\nu_2}^\perp \right )\!.~~~
\ee
For higher states, the operator can be calculated using the
recurrent expression:
\be &&O^{\mu_1\ldots\mu_n}_{\nu_1\ldots
\nu_n}=\frac{1}{n^2} \bigg (
\sum\limits_{i,j=1}^{n}g^\perp_{\mu_i\nu_j}
O^{\mu_1\ldots\mu_{i-1}\mu_{i+1}\ldots\mu_n}_{\nu_1\ldots
\nu_{j-1}\nu_{j+1}\ldots\nu_n}
\nonumber \\
 &&  -\frac{4}{(2n-1)(2n-3)}
\nn    && \times\sum\limits_{i<j\atop k<m}^{n}
g^\perp_{\mu_i\mu_j}g^\perp_{\nu_k\nu_m}
O^{\mu_1\ldots\mu_{i-1}\mu_{i+1}\ldots\mu_{j-1}\mu_{j+1}\ldots\mu_n}_
{\nu_1\ldots\nu_{k-1}\nu_{k+1}\ldots\nu_{m-1}\nu_{m+1}\ldots\nu_n}
\bigg )\,.
\ee
The tensor $F^{\mu_1\ldots\mu_n}_{\nu_1\ldots \nu_n}$ has all
orthogonality properties of the tensor
$O^{\mu_1\ldots\mu_n}_{\nu_1\ldots \nu_n}$ plus orthogonality to the
$\gamma$-matrix:
\be
\gamma_{\mu_i}F^{\mu_1\ldots\mu_n}_{\nu_1\ldots \nu_n}
=F^{\mu_1\ldots\mu_n}_{\nu_1\ldots \nu_n}\gamma_{\nu_j}=0\;.
\ee

The pseudoscalar meson-nucleon vertices for the partial wave with
spin $J$ have the form:
\be
\label{piN_vertex}
Q^{(+)}_{\mu_1\ldots\mu_n}&=&
X^{(n)}_{\mu_1\ldots\mu_n}(q^\perp)u(q_1)\,, \nn
Q^{(-)}_{\mu_1\ldots\mu_n}&=& i\gamma_5 \gamma_\nu
X^{(n+1)}_{\nu\mu_1\ldots\mu_{n}}(q^\perp)u(q_1) \, ,
\ee
where $n=J\!-\!1/2$ and $u(q_1)$ is the bispinor of the baryon. The
'+' and '-' indices describe two sets of the partial waves with
relation between orbital momentum $L$ and the total spin $J$ as
$J=L+1/2$ ('+' partial waves) and $J=L-1/2$ ('-' partial waves). The
'+' set of vertices describes the partial waves with
$J^P=\frac{1}{2}^-$, $\frac{3}{2}^+$, $\frac{5}{2}^-\ldots$ and the
second set $J^P=\frac{1}{2}^+$, $\frac{3}{2}^-$,
$\frac{5}{2}^+\ldots$.

In the case of virtual photons there are, for every partial wave
with $J>1/2$, three $\gamma^* N$ vertices; for real photons, only
two of them are independent \cite{Anisovich:2006bc}. In the $LS$
formalism these vertices correspond to spin $1/2$ and $3/2$ of the
photon-nucleon system. For '+' states the vertices are (following
the ordering in \cite{Anisovich:2006bc}):
\bea
\label{vf_plus}
Q^{(1+)}_{\alpha_1\ldots\alpha_n}&=& \bar u(k_1)\gamma^\perp_\mu
i\gamma_5
X^{(n)}_{\alpha_1\ldots\alpha_n}(k^\perp)\varepsilon_\mu\,, \nn
Q^{(3+)}_{\alpha_1\ldots\alpha_n}&=&\bar u(k_1) \gamma_\nu i
\gamma_5 X^{(n)}_{\nu\alpha_1\ldots\alpha_{n-1}}(k^\perp)
g^\perp_{\mu\alpha_n}\varepsilon_\mu \,,
\eea
where $\bar u(k_1)$ is bispinor of the initial nucleon and
$\varepsilon_\mu$ is the photon polarization vector.

For '-' states we have:
\bea
\label{vf_minus}
Q^{(1-)}_{\alpha_1\ldots\alpha_n}&=&\bar u(k_1)
\gamma_\xi\gamma^\perp_\mu\varepsilon_\mu
X^{(n+1)}_{\xi\alpha_1\ldots\alpha_{n}}(k^\perp) \;, \nn
Q^{(3-)}_{\alpha_1\ldots\alpha_{n}}&=&\bar u(k_1)
X^{(n-1)}_{\alpha_2\ldots\alpha_{n}}(k^\perp)
g^\perp_{\alpha_1\mu}\varepsilon_\mu \;.
\eea

The orbital angular momentum operators for $L \le 3 $ are:
\bea
X^{(0)}&=&1\ , \qquad X^{(1)}_\mu=k^\perp_\mu\ , \qquad
\nonumber \\
X^{(2)}_{\mu_1 \mu_2}&=&\frac32\left(k^\perp_{\mu_1}
k^\perp_{\mu_2}-\frac13\, k^2_\perp g^\perp_{\mu_1\mu_2}\right),
\nonumber  \\
X^{(3)}_{\mu_1\mu_2\mu_3}&=&\frac52\Big[k^\perp_{\mu_1} k^\perp_{\mu_2 }
k^\perp_{\mu_3}
\nn
&-&
\frac{k^2_\perp}{5}
\left(g^\perp_{\mu_1\mu_2}k^\perp_{\mu_3}+g^\perp_{\mu_1\mu_3}k^\perp_{\mu_2}+
g^\perp_{\mu_2\mu_3}k^\perp_{\mu_1}
\right)\Big]\,.~~~
\eea

The operator $X^{(n+1)}_{\nu\mu_1\ldots\mu_n}$
can be written as a series of products of metric tensors and
relative momentum vectors. The first term is proportional to the
production of relative momentum vectors $k^\perp_\mu$, other terms
correspond to the substitution of two vectors by a metric
tensor with corresponding indices:
\bea
&&X^{(n+1)}_{\nu\mu_1\ldots\mu_n}(k^\perp) =\alpha_{n+1} \bigg [
k^\perp_{\nu}k^\perp_{\mu_1}k^\perp_{\mu_2}k^\perp_{\mu_3}
\ldots k^\perp_{\mu_n}-
\frac{k^2_\perp}{2n\!+\!1}
\nn
&&\times\bigg(\sum\limits_{i=1}^n
g^\perp_{\nu\mu_i}\prod\limits_{j\ne i} k^\perp_{\mu_j}+
\sum\limits_{i<j}^n
g^\perp_{\mu_i\mu_j}k^\perp_{\nu}
\!\!\prod\limits_{m\ne i\ne j}\!\!
k^\perp_{\mu_m}+\ldots \bigg )
\nn
&&+\frac{k^4_\perp}{(2n\!+\!1)(2n\!-\!1)}\bigg(
\sum\limits_{i,j<m}^n
g^\perp_{\nu\mu_i}g^\perp_{\mu_j\mu_m}
\!\!\prod\limits_{l\ne i\ne j\ne m}\!\! k^\perp_{\mu_l}
\nn
&&+\sum\limits_{i<k,j<m}^n
g^\perp_{\mu_i\mu_k}g^\perp_{\mu_j\mu_m} k^\perp_{\nu}
\prod\limits_{^{l\ne i\ne k}_{\ne j\ne m}} k^\perp_{\mu_l}+
\ldots\bigg)+\ldots\bigg ].~~~~~~~~
\label{x-direct}
\eea

\section*{Appendix B: Contribution of the loop diagrams}

For the $\pi N$ vertices we have \cite{Anisovich:2006bc}:
\bea
W^{(+)}_n&=&(-1)^n \frac{\alpha_n}{2n+1}|\vec k|^{2n}
\frac{m_N+k_{10}}{2m_N}\;,
 \nn
W^{(-)}_n&=&(-1)^n \frac{\alpha_{n+1}}{n+1} |\vec k|^{2n+2}
\frac{m_N+k_{10}}{2m_N}\;,
\label{w_piN_minus}
\eea

\begin{figure}
\centerline{\epsfig{file=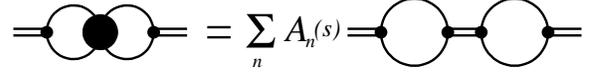,width=8cm}} \caption{Diagram
representation of eq. (\ref{decomp_2})}
\label{fig::scalor_proj}
\end{figure}

and for the $\gamma N$ vertices (in the case of photoproduction):
\bea
W^{(11+)}_n&=&(-1)^n2\frac{\alpha_n}{2n+1}|\vec k|^{2n}
\frac{m_N\!+\!k_{10}}{2m_N}\;,
 \nn
W^{(33+)}_n&=&(-1)^n\frac{\alpha_n}{2n+1}\frac{(n+1)}{n} |\vec
k|^{2n}\frac{m_N\!+\!k_{10}}{2m_N}\;,
 \nn
W^{(13+)}_n&=&(-1)^n\frac{\alpha_n}{2n+1}|\vec
k|^{2n}\frac{m_N\!+\!k_{10}}{2m_N}\,.
\label{w_plus}
\eea
for the '+' states and
\bea
W^{(11-)}_n&=&(-1)^n\frac{2\alpha_{n+1}}{n+1}|\vec k|^{2n+2}
\frac{m_N\!+\!k_{10}}{2m_N}\;,
 \nn
W^{(33-)}_n&=&(-1)^n\frac{\alpha_{n-1}(n+1)}{(2n\!+\!1)(2n\!-\!1)}
|\vec k|^{2n-2}\frac{m_N\!+\!k_{10}}{2m_N}\;, \nn
W^{(13-)}_n&=&(-1)^n\frac{\alpha_{n-1}}{n+1}|\vec k|^{2n}
\frac{m_N\!+\!k_{10}}{2m_N}\,
\label{w_minus}
\eea
for the '-' states.
 The $\gamma N$ vertices in this representation
are not orthogonal to each other, and to extract partial waves one
needs to solve a $2\times2$ system of linear equations.

\section*{Appendix C: Single meson photoproduction amplitude}

The general structure of the single--meson photoproduction amplitude
in c.m.s. of the reaction is given by
\bea
J_\mu\!=\! i {\mathcal F_1}
 \sigma_\mu\! +&&\!\!{\mathcal F_2} (\vec \sigma \vec q)
\frac{\varepsilon_{\mu i j} \sigma_i k_j}{|\vec k| |\vec q|} \!+\!i
{\mathcal F_3} \frac{(\vec \sigma \vec k)}{|\vec k| |\vec q|} q_\mu
\!+\!i {\mathcal F_4} \frac{(\vec \sigma \vec q)}{\vec q^2} q_\mu\,,
\nn  &&A=\omega^*J_\mu\varepsilon_\mu \omega'\,,
\label{amp_t_gammaN_cms}
\eea
where $\vec q$ is the momentum of the nucleon in the $\pi N$ channel
and $\vec k$ the momentum of the nucleon in the $\gamma N$ channel
calculated in  the c.m.s. of the reaction. The $\sigma_i$ are Pauli
matrices and $\omega$, $\omega'$ are non relativistic spinors of
initial and final states correspondingly.

If ${\mathcal F_i}$ are known, e.g. from the $t$ or $u$ channel
exchange amplitudes calculated in the c.m.s. of the reaction, the
partial wave amplitudes can be obtained as
\be
A^{(i\pm)}_n&=&\int\limits_{-1}^{1} \frac{dz}{2}{\mathcal
F_m}D^{(i\pm)}_m\,,
\ee
where $z$ is the cosine of the angle between initial and final
relative momenta and vectors $D^{(i\pm)}$ are equal to
\be
D^{(1+)}&=&\frac{1}{\kappa_n\alpha_n}
\left(P_n,-P_{n+1},0,\frac{(1\!-\!z^2)
P'_{n+1}}{(n\!+\!1)(n\!+\!2)}\right)\ ,
\nn
D^{(2+)}&=&\frac{1-z^2}{\kappa_n\alpha_n}
\left(0,0,\frac{P'_n}{(n\!+\!1)}, \frac{n
P'_{n+1}}{(n\!+\!1)(n\!+\!2)}\right )\ ,
\nn
D^{(1-)}&=&-\frac{n\!+\!1}{\kappa_{n+1}\alpha_{n+1}}
\left(-P_{n+1},P_n,\frac{(1\!-\!z^2)P'_{n+1}}{(n\!+\!1)(n\!+\!2),0}\right)\ ,
\nn
D^{(2-)}&=&-\frac{1-z^2}{\kappa_{n-1}\alpha_{n-1}|\vec
k|^2}\left (0,0,\frac{nP'_{n+1}}{(n\!+\!2)}, P'_n\right ).
\label{amp_dec}
\ee
Here $P_n=P_n(z)$ are Legendre polynomials and $P'_n=dP_n(z)/dz$.

Using the multipole decomposition of the $A^{(i\pm)}_n$ amplitudes
given in \cite{Anisovich:2004zz} one can obtain the standard
expression for the projection of the total amplitude into multipoles.

\section*{Appendix D: Reggeon propagator parametrization }

In this section we give the expressions for Reggeon propagators used in the fit.

The propagator for pion exchange has the form
\be
R_{\pi}(+,\nu,t)=\frac{e^{-i\frac{\pi}{2}\alpha_{\pi}(t)}}
{\sin (\frac{\pi}{2}\alpha_{\pi}(t)) \Gamma \left (\frac{
\alpha_{\pi}(t)}{2} +1\right )}
\left (\frac{\nu}{\nu_0}\right )^{\alpha_{\pi}(t)}\;,\quad
\ee
where $\alpha_{\pi}(t)=-0.014+0.72t$ is a function defining the
trajectory, $\nu_0$ is a normalization factor (which can be taken to
be 1). The $\Gamma$-function is introduced in the denominator to
eliminate the additional poles at $t<0$. The propagator for Kaon
exchange is given by
\be
R_{\rm K}(+,\nu,t)=\frac{e^{-i\frac{\pi}{2}\alpha_{\rm K}(t)}}
{\sin (\frac{\pi}{2}\alpha_{\rm K}(t)) \Gamma \left (\frac{
\alpha_{\rm K}(t)}{2} +1\right )}
\left (\frac{\nu}{\nu_0}\right )^{\alpha_{\rm K}(t)}\;,\quad
\ee
where $\alpha_{\rm K}(t)=-0.25+0.85t$.

The propagator
for $\rm K^*$ exchange is identical to the $\rho$ exchange propagator but has $\alpha_{\rm
K^*}(t)=0.32+0.85t$.

\end{document}